\pdfoutput=1
\documentclass[
reprint,
showpacs,preprintnumbers,
nofootinbib,
amsmath,amssymb,
aps,
pra,
]{revtex4-1}

\usepackage{graphicx}
\usepackage{dcolumn}
\usepackage{bm}
\usepackage{multirow}
\usepackage{subcaption}

\begin{document}

\title{Generalized isospin, generalized mass groups, and generalized Gell-Mann--Okubo formalism}

\author{N. Beaudoin}
\affiliation{Universit\'e de Moncton, Moncton, NB, E1A 3E9, Canada}
\author{G. Landry}
\email[To whom correspondence should be addressed: ]{g.landry@dal.ca}
\affiliation{Universit\'e de Moncton, Moncton, NB, E1A 3E9, Canada}
\affiliation{Dalhousie University -- Agricultural Campus, Truro, NS, B2N 5E3, Canada}
\author{R. Sandapen}
\affiliation{Universit\'e de Moncton, Moncton, NB, E1A 3E9, Canada}
\affiliation{Mount Allison University, Sackville, NB, E46 1E6, Canada}

\date{\today}

\begin{abstract}
The current concepts of isospin and baryon mass groups are only well-adapted to deal with baryon multiplets involving both the $u$ and $d$ quarks, and some other quark $k$. In this paper, we generalize isospin and mass groups to accommodate baryon multiplets involving quarks of any flavor, and the Gell-Mann--Okubo (GMO) formalism is generalized accordingly. Generalized isospin proves to be a simple and valuable framework when working in non-$udk$ baryon multiplets, and provides new quantum numbers that allows us to distinguish $\Lambda$-like baryons from $\Sigma$-like baryons in the non-$udk$ multiplets. The generalized GMO formalism allows us to quantify the quality of flavor symmetries seen in baryon multiplets, and also allows us to predict the masses of all observable $J^P = \frac{1}{2}^{+}$ and $\frac{3}{2}^{+}$\linebreak baryons with an estimated accuracy on the order of 50~MeV in the worst cases, on mass scales that span anywhere from 1000~MeV to 15000~MeV.

\end{abstract}


\maketitle

\section{Introduction}
\begin{figure*}[!t]
 \centering
 \begin{subfigure}{0.45\textwidth}
 \centering
 \includegraphics[width=\textwidth]{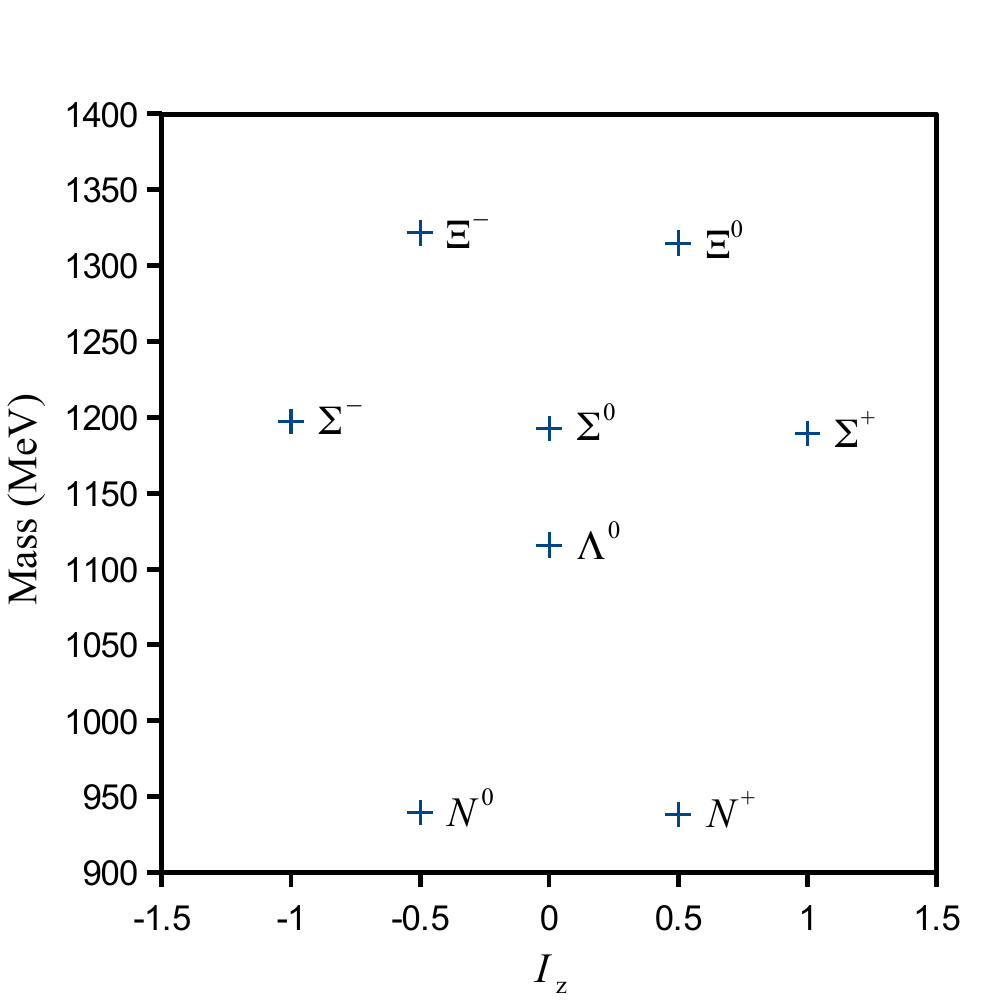}
 \caption{\label{Fig-1a}The light baryon octet.}
 \end{subfigure}
 \begin{subfigure}{0.45\textwidth}
 \centering
 \includegraphics[width=\textwidth]{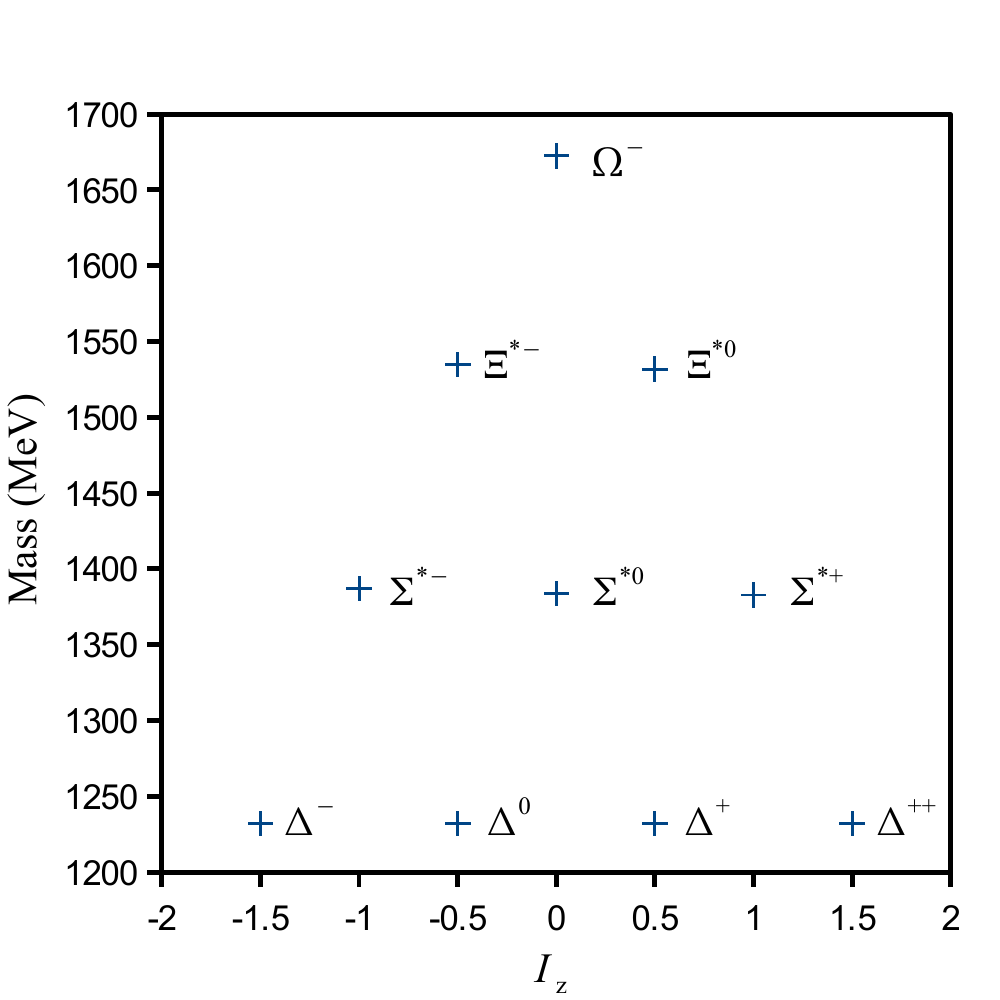}
 \caption{\label{Fig-1b}The light baryon decuplet.}
 \end{subfigure}

 \begin{subfigure}{0.45\textwidth}
 \centering
 \includegraphics[width=\textwidth]{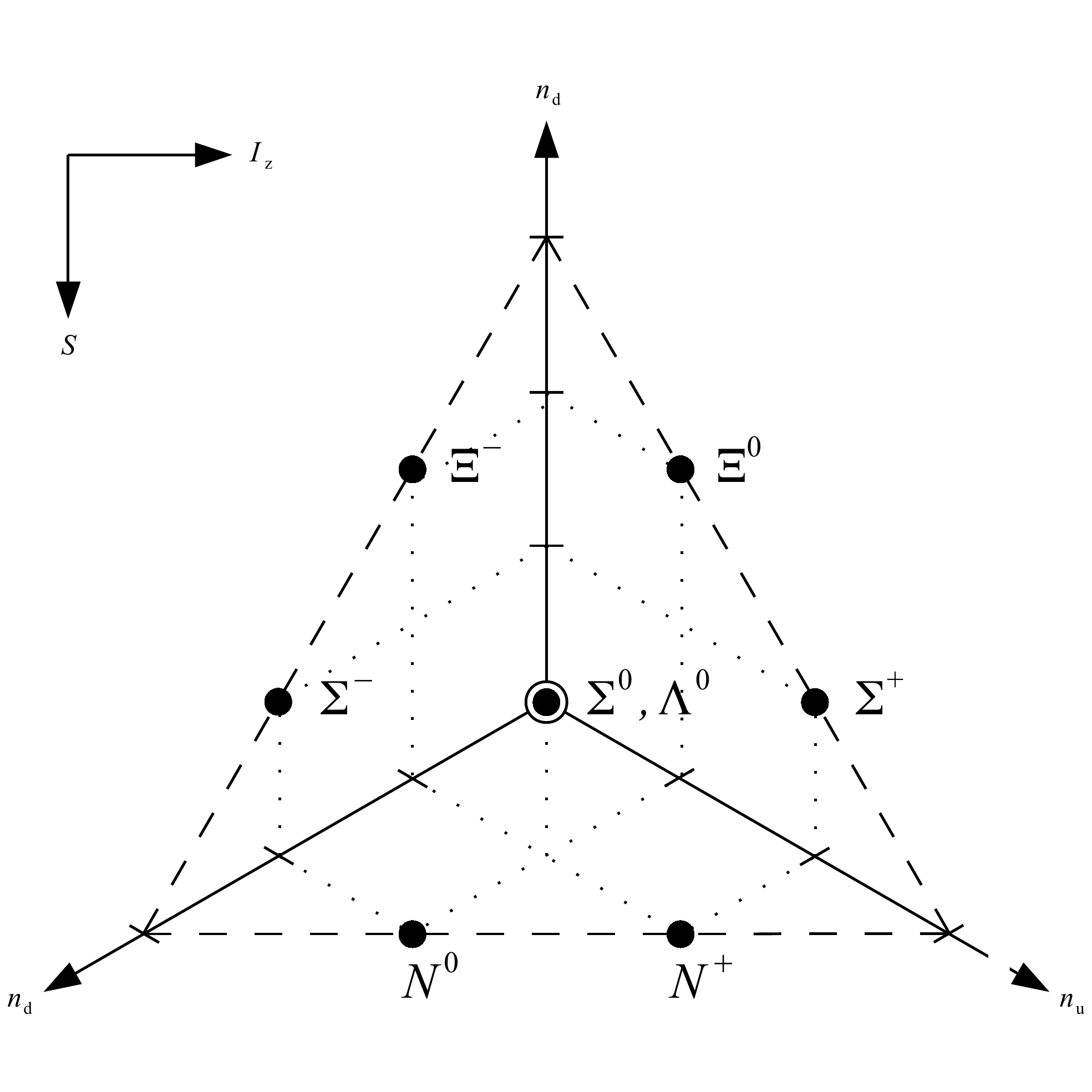}
 \caption{\label{Fig-1c}SU(3) weight diagrams for the light baryon octet.}
 \end{subfigure} 
 \begin{subfigure}{0.45\textwidth}
 \centering
 \includegraphics[width=\textwidth]{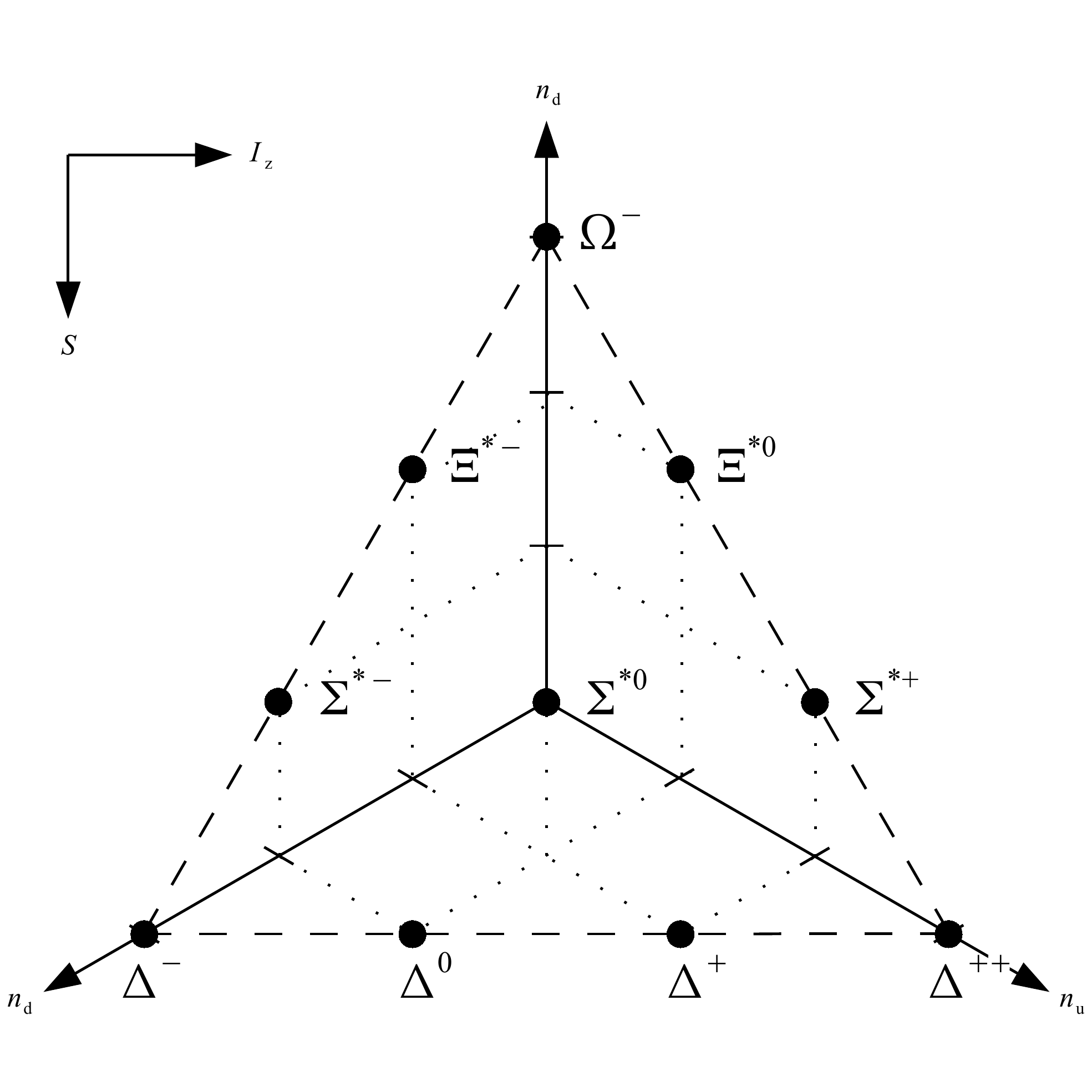}
 \caption{\label{Fig-1d}SU(3) weight diagrams for the light baryon decuplet.}
 \end{subfigure}
\caption{\label{Fig-1}Mass vs. $I_\mathrm{z}$ and SU(3) weight diagrams for the light baryon multiplets. The quark model explained $I_\mathrm{z}$ in terms of the $u$ and $d$ quarks [Eq.~(\protect\ref{eq-8})], and $S$ in terms of the $s$ quark [Eq.~(\protect\ref{eq-9})]. Masses are taken from \protect\cite{PDG2012}. In the number of quark space, baryon states lie on a plane which intersects the axes at (3,0,0), (0,3,0), and (0,0,3). Note that the $\Sigma^0$, $\Lambda^0$, and $\Sigma^{*0}$ baryons lie at (1,1,1) and not at (0,0,0).}
\end{figure*}

The concepts of isospin, proposed by Heisenberg in 1932 \cite{Heisenberg1932a, Heisenberg1932b, Heisenberg1933} to explain the similar masses of nucleons, and strangeness, following the efforts of Nakano, Nishijima and Gell-Mann in the mid-1950s \cite{Nakano1953, Nishijima1955, Gellmann1956} to explain decay properties of particles such as the $\Sigma$ baryons and $K$ mesons, are of key importance in hadron physics. These efforts culminated in the Gell-Mann--Nishijima (GMN) formula for the charge of hadrons:
\begin{equation}\label{eq-1}
Q = I_\mathrm{z} + \frac{1}{2}\left( B' + S \right),
\end{equation}
and the multiplicity relation:
\begin{equation}\label{eq-2}
\mathrm{mult}\left(I_\mathrm{z}\right) = 2I+1,
\end{equation}
where $Q$ is the charge number, $I$ is isospin, $I_\mathrm{z}$ is the isospin projection, $B'$ is the baryon number, and $S$ is strangeness. The $I$ and $S$ values of the light baryon mass groups are summarized in Table~\ref{tab-1}. This paved the way for Gell-Mann \cite{Gellmann1961} and Ne'emann \cite{Neemann1961} to propose the Eightfold Way in the early 1960s. The Eightfold Way explained the patterns observed in mass~vs.~$I_\mathrm{z}$ diagrams in terms of a broken SU(3) symmetry (see Fig.~\ref{Fig-1}). In particular, it allowed Gell-Mann \cite{Gellmann1961} and Okubo \cite{Okubo1962a, Okubo1962b} to develop a mass formula for hadrons, known as the\linebreak Gell-Mann--Okubo (GMO) formula:\footnote{A concise derivation of the GMO formula is also available in \protect\cite{Goldberg1963}.}
\begin{equation}\label{eq-3}
M\left( I, S \right) = a_0 - a_1 S + a_2 \left[ I \left( I+1 \right) - \frac{1}{4} S^2 \right] ,
\end{equation}
where $a_0$, $a_1$, and $a_2$ are free parameters\footnote{The parameters are chosen so that they are positive.} specific to a given multiplet. 

\begin{table}[b]
\caption{\label{tab-1}Light baryon mass groups}
\renewcommand{\arraystretch}{1.4}
\begin{ruledtabular}
\begin{tabular}{cccc}
Multiplet & Mass group\footnote{For notational convenience, we use $\Sigma^*$, $\Xi^*$, and $\Lambda^\dagger$ to refer to the $\Sigma(1385)$, $\Xi(1530)$, and $\Lambda(1405)$ groups, respectively, as well as their counterparts in higher mass multiplets.} & $I$ & $S$ \\ 
\hline
\multirow{4}{*}{Octet} & $N^{\phantom{*}}$ & 1/2 & $\phantom{-}0$ \\ 
 & $\Lambda^{\phantom{*}}$ & 0 & $-1$ \\ 
 & $\Sigma^{\phantom{*}}$ & 1 & $-1$ \\ 
 & $\Xi^{\phantom{*}}$ & 1/2 & $-2$ \\ 
\hline
\multirow{4}{*}{Decuplet} & $\Delta^{\phantom{*}}$ & 3/2 & $\phantom{-}0$ \\ 
 & $\Sigma^{*}$ & 1 & $-1$ \\ 
 & $\Xi^{*}$ & 1/2 & $-2$ \\ 
 & $\Omega^{\phantom{*}}$ & 0 & $-3$ \\ 
\hline
Singlet & $\Lambda^{\dagger}$ & 0 & $-1$ \\ 
\end{tabular}
\end{ruledtabular}
\end{table}

\begin{figure*}[!t]
 \centering
 \begin{subfigure}{0.475\textwidth}
 \centering
 \includegraphics[width=\textwidth]{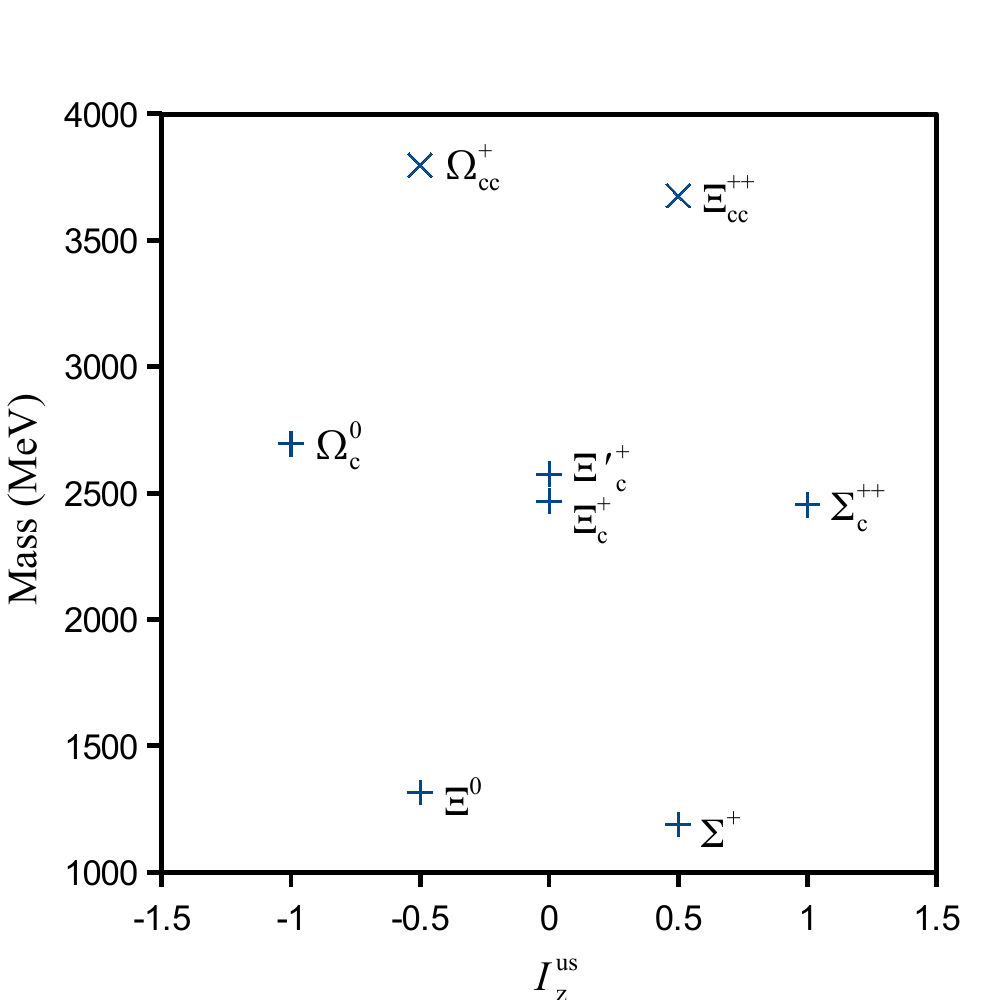}
 \caption{\label{Fig-2a}The $usc$ octet.}
 \end{subfigure}
 \begin{subfigure}{0.475\textwidth}
 \centering
 \includegraphics[width=\textwidth]{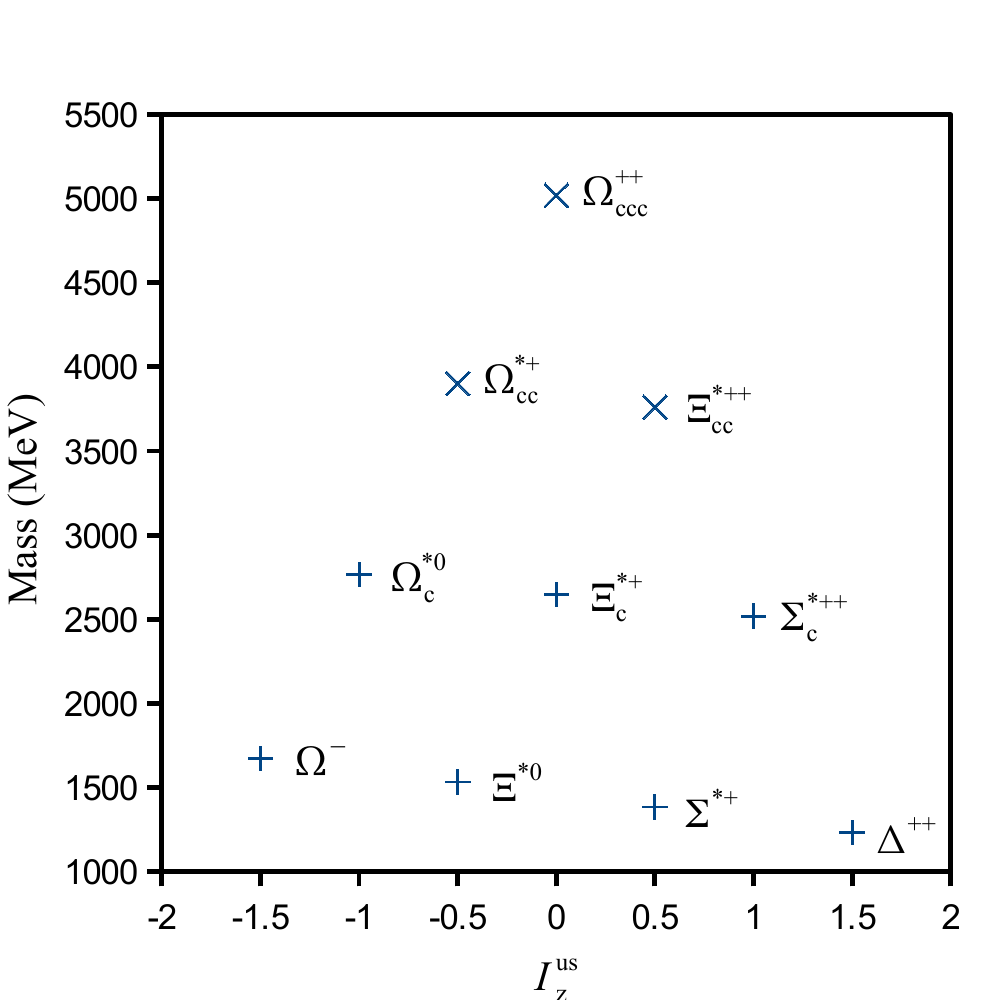}
 \caption{\label{Fig-2b}The $usc$ decuplet.}
 \end{subfigure}
\caption{\label{Fig-2}The $usc$ multiplets. Measured masses are taken from \protect\cite{PDG2012} and are indicated by +, while masses predicted using Eq.~(\protect\ref{eq-19}) and parameter values from Table~\protect\ref{tab-4} and Table~\protect\ref{tab-5} are indicated by $\times$. $I_\mathrm{z}^{us}$ is the analogue of $I_\mathrm{z}$ in this multiplet [see Eq.~(\protect\ref{eq-17})].}
\end{figure*}

The GMO formula is more commonly encountered as\footnote{For convenience, we use the mass group symbols to denote the average mass of the mass groups.}
\begin{equation}\label{eq-4}
\frac{N+\Xi}{2} = \frac{3\Lambda + \Sigma}{4}
\end{equation}
for the light baryon octet, and as the equal-spacing rule
\begin{equation}\label{eq-5}
\Omega - \Xi^* = \Xi^* - \Sigma^* = \Sigma^* - \Delta = a_1 - 2a_2
\end{equation}
for the light baryon decuplet. We note that a similar relation exists in the octet as well:
\begin{equation}\label{eq-6}
\Xi - \Sigma = a_1 - 2a_2,
\end{equation}
and we also note the following relation:
\begin{equation}\label{eq-split-1}
\Sigma - \Lambda = 2a_2.
\end{equation}

The GMO formula reproduces the masses of light baryons with a root-mean-square (RMS) error of $7$~MeV in the octet and $3$~MeV in the decuplet, and famously allowed Gell-Mann to predict the existence and mass of the $\Omega^-$ baryon \cite{Gellmann1962, Barnes1964}, based on the equal-spacing rule [Eq.~(\ref{eq-5})], firmly establishing the validity of the Eightfold Way.

Shortly after the discovery of the $\Omega^-$, Gell-Mann proposed the quark model, which provides the physical basis behind the Eightfold Way. In particular,\linebreak the baryon number, the isospin projection, and strangeness are explained in terms of\linebreak the numbers of $u$, $d$, and $s$ quarks (see Fig.~\ref{Fig-1}):
\begin{align}
S & = -n_\mathrm{s}, \label{eq-7} \\ 
I_\mathrm{z} & = \frac{1}{2} \left( n_\mathrm{u} - n_\mathrm{d} \right), \label{eq-8}\\ 
B' & = \frac{1}{3} \left( n_\mathrm{u} + n_\mathrm{d} + n_\mathrm{s} \right).\label{eq-9}
\end{align}
These, when substituted in the GMN formula [Eq.~(\ref{eq-1})], would yield
\begin{equation}\label{eq-10}
Q = +\frac{2}{3}\left( n_\mathrm{u} \right) - \frac{1}{3}\left( n_\mathrm{d} + n_\mathrm{s} \right),
\end{equation}
i.e., the charge of a hadron is simply due to the charge of its constituent quarks. The equal-spacing rule can also be understood as a consequence of the $u$ and $d$ masses being very similar to each other, with the $s$ mass being higher:
\begin{equation}\label{eq-11}
a_1 - 2a_2 = m_\mathrm{s} - \frac{1}{2}\left(m_\mathrm{u} + m_\mathrm{d} \right).
\end{equation}
That is, with each increase in $n_s$, one adds one $s$ quark, and removes either one $u$ or $d$ quark. Therefore the mass accordingly increases by $m_\mathrm{s}$, and decreases (on average) by the average of $m_\mathrm{u}$ and $m_\mathrm{d}$, where $m_u$, $m_d$, and $m_s$ are the bare masses of the $u$, $d$, and $s$ quarks, respectively.\footnote{In the decuplet, the mass of a baryon can be said to be due to the bare mass of its valence quarks plus some interaction term. Within the decuplet, this interaction term should be constant for all members, as all members share the same symmetries. Evaluating the mass difference between mass group therefore probes the bare mass of quarks.}

In terms of representation theory, flavor symmetries in light baryons can be described as ${\mathbf{3}\otimes\mathbf{3}\otimes\mathbf{3} = \mathbf{10} \oplus \mathbf{8} \oplus \mathbf{8} \oplus \mathbf{1}}$, with the baryon decuplet, octet, and singlet being associated with $\mathbf{10}$, $\mathbf{8}$, and $\mathbf{1}$ respectively.

The additional concepts introduced by the discovery of the $c$, $b$, and $t$ quarks are simply the introduction of three new flavor quantum numbers:
\begin{align}
C = +n_\mathrm{c}, \label{eq-12} \\ 
B = -n_\mathrm{b}, \label{eq-13} \\ 
T = +n_\mathrm{t}, \label{eq-14}
\end{align}
the generalization of baryon number [Eq.~(\ref{eq-9})] to
\begin{equation}\label{eq-15}
B\ = \frac{1}{3} \left( n_\mathrm{u} + n_\mathrm{d} + n_\mathrm{s} + n_\mathrm{c} + n_\mathrm{b} + n_\mathrm{t}\right),
\end{equation}
and the generalization of the charge formula [Eqs.~(\ref{eq-1}) and (\ref{eq-10})] to respectively
\begin{subequations}
\begin{align}
Q & = I_\mathrm{z} + \frac{1}{2}\left( B'+S+C+B+T \right), \label{eq-16a}\\
{} & = +\frac{2}{3}\left( n_\mathrm{u} + n_\mathrm{c} + n_\mathrm{t} \right) - \frac{1}{3}\left( n_\mathrm{d} + n_\mathrm{s} + n_\mathrm{b} \right). \label{eq-16b}
\end{align}
\end{subequations}
In terms of representation theory, flavor symmetries in baryons are described as ${\mathbf{6}\otimes\mathbf{6}\otimes\mathbf{6} = \mathbf{56} \oplus \mathbf{70} \oplus \mathbf{70} \oplus \mathbf{20}}$. Unfortunately, this is usually all that is said concerning the additional baryons, with the usual caveat that SU($N$) symmetries get worse as $N$ increases. For instance, while one could select any three flavors of quark study the symmetries of the associated multiplets on their own (see Fig.~\ref{Fig-2} for the $usc$ multiplets) there is a lack of obvious framework in which to do so, and one must usually resort to \textit{ad hoc} analogies in those cases. For example, Mahanthappa and Unger talked of ``$U$-spin''(isospin analogue for $d$ and $s$ symmetries) and ``$L$-spin'' (isospin analogue for $u$ and $c$ symmetries) in \cite{Mahanthappa1977}, to study baryons in the context of SU(4) symmetries.

\section{Generalized isospin and generalized mass groups}

\begin{figure}[!t]
 \centering
 \includegraphics[width=0.45\textwidth]{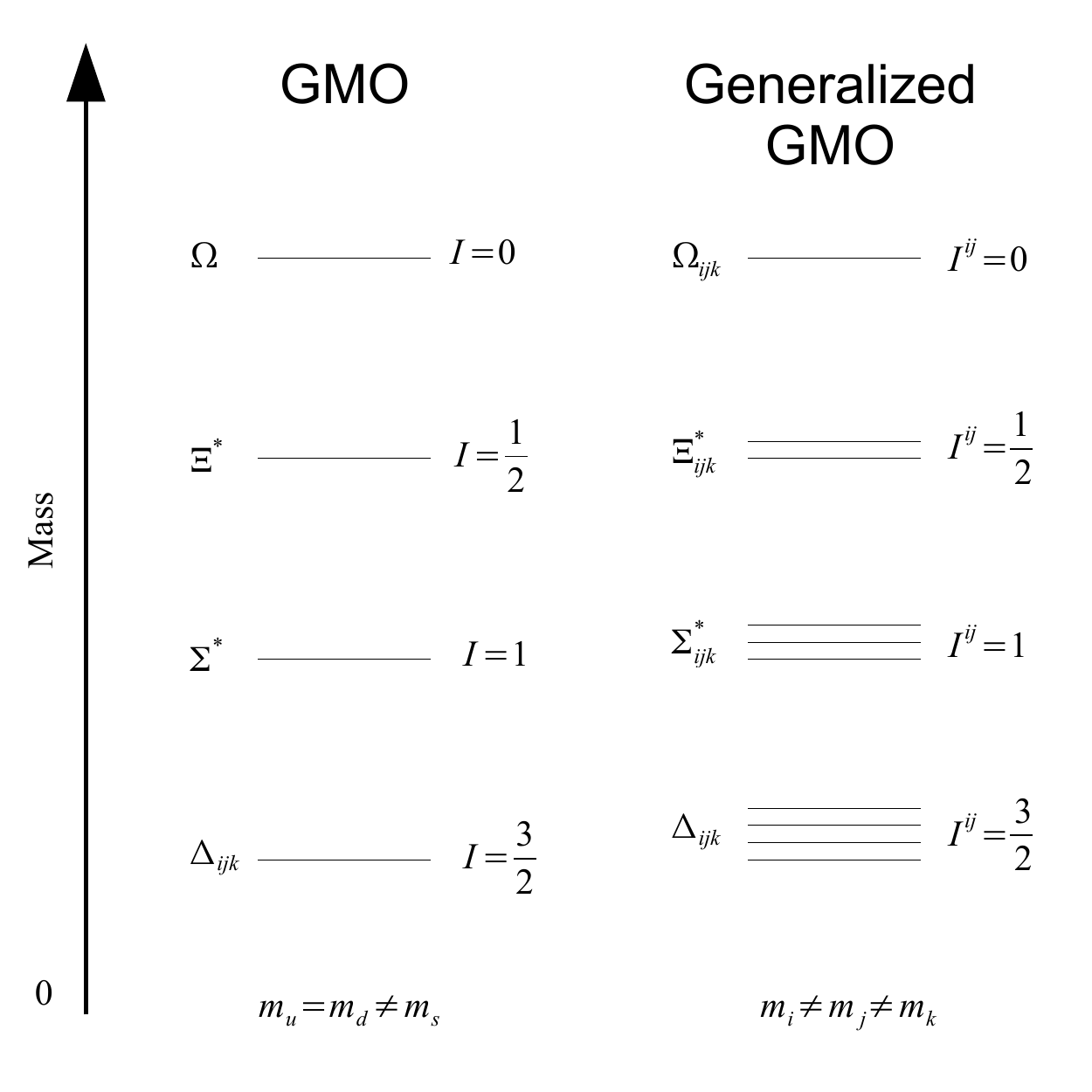}
 \caption{\label{Fig-symbreak}In the GMO formalism (left), the $\Delta$, $\Sigma^*$, $\Xi^*$, and $\Omega$ mass groups are degenerate. In the generalized GMO formalism (right), the degeneracy of the $\Delta_{ijk}$, $\Sigma^*_{ijk}$, $\Xi^*_{ijk}$, and $\Omega_{ijk}$ mass groups is lifted.}
\end{figure}

We therefore introduce the concept of generalized isospin\footnote{The associated SU(3) operators can be found in Appendix~\ref{Appendix}} in a given multiplet of $ijk$ flavors (with ${i \neq j \neq k \in \left\{\mathrm{u, d, s, c, b, t} \right\}}$), by analogy with ``ordinary'' isospin [Eqs.~(\ref{eq-8}) and (\ref{eq-2})]:
\begin{align}
I^{ij}_\mathrm{z} & = \frac{1}{2}\left( n_i-n_j \right), \label{eq-17} \\ 
\mathrm{mult}\left( I^{ij}_\mathrm{z} \right) & = 2I^{ij}+1, \label{eq-18}
\end{align}
and we introduce the concept of generalized mass groups (Table~\ref{tab-2}) by analogy with the light baryon mass groups (Table~\ref{tab-1}). This easily allows us to generalize the GMO formula [Eq.~(\ref{eq-3})] to \textit{any} baryon multiplet. If we set ${m_k > m_j = m_i}$, we would obtain the exact same formula as before. However, by instead setting ${m_k > m_j > m_i}$, the mass degeneracy of the generalized isomultiplets is lifted (see Fig.~\ref{Fig-symbreak}), and we instead obtain
\begin{equation}\label{eq-19}
\begin{split}
M \left( I^{ij}, I^{ij}_\mathrm{z}, n_k \right) = & ~a^{ijk}_0 + a^{ijk}_1 n_k \\ 
{} & + a^{ijk}_2 \left[ I^{ij}\left( I^{ij}+1 \right)-\frac{1}{4}n_k^2 \right] \\ 
{} & - a^{ijk}_3 I_\mathrm{z}^{ij},
\end{split}
\end{equation}
where $a^{ijk}_0$, $a^{ijk}_1$, $a^{ijk}_2$, and $a^{ijk}_3$ are free parameters\footnote{The parameters are chosen so that they will be positive when we choose $ijk$ so that ${m_k > m_j > m_i}$.} specific to a given baryon multiplet involving the $ijk$ flavours.\footnote{To retrieve the original GMO results, one only has to set $i = \mathrm{u}$, $j = \mathrm{d}$, $k = \mathrm{s}$, and $a_3^{ijk} = 0$.} The last term is introduced to account for the ``skewedness'' of the patterns found in mass vs. $I^{ij}_\mathrm{z}$ diagrams (see Fig.~\ref{Fig-2} for example), and will be justified later in this section.

\begin{table}[!]
\caption{\label{tab-2}Generalized mass groups}
\renewcommand{\arraystretch}{1.4}
\begin{ruledtabular}
\begin{tabular}{cccc}
Multiplet & Mass group & $I^{ij}$ & $n_k$ \\ 
\hline
\multirow{4}{*}{Octets} & $N_{ijk}$ & 1/2 & 0 \\ 
 & $\Lambda_{ijk}$ & 0 & 1 \\ 
 & $\Sigma_{ijk}$ & 1 & 1 \\ 
 & $\Xi_{ijk}$ & 1/2 & 2 \\ 
\hline
\multirow{4}{*}{Decuplets} & $\Delta_{ijk}$ & 3/2 & 0 \\ 
 & $\Sigma^{*}_{ijk}$ & 1 & 1 \\ 
 & $\Xi^{*}_{ijk}$ & 1/2 & 2 \\ 
 & $\Omega_{ijk}$ & 0 & 3 \\ 
\hline
Singlet & $\Lambda^{\dagger}_{ijk}$ & 0 & 1 \\ 
\end{tabular}
\end{ruledtabular}
\end{table}

These generalizations will yield the same type of mass relations as before. For octets, using Eq.~(\ref{eq-19}) and values from Table~\ref{tab-2} leads to\footnote{Here again, we use the mass group symbols to denote the average mass of mass groups.}
\begin{equation}\label{eq-20}
\frac{N_{ijk} + \Xi_{ijk}}{2} = \frac{3\Lambda_{ijk} + \Sigma_{ijk}}{4},
\end{equation}
where we again have
\begin{equation}\label{eq-21}
\Xi_{ijk} - \Sigma_{ijk} = a^{ijk}_1 - 2a^{ijk}_2,
\end{equation}
and to
\begin{equation}\label{eq-split-2}
\Sigma_{ijk} - \Lambda_{ijk} = 2a^{ijk}_2.
\end{equation}
For decuplets, using Eq.~(\ref{eq-19}) and values from Table~\ref{tab-2} again leads to equal-spacing rules:
\begin{equation}\label{eq-22}
\Omega_{ijk} - \Xi^*_{ijk} = \Xi^*_{ijk} - \Sigma^*_{ijk} = \Sigma^*_{ijk} - \Delta_{ijk} = a^{ijk}_1 - 2a^{ijk}_2.
\end{equation}
Again we note that the $a^{ijk}_0$, $a^{ijk}_1$, $a^{ijk}_2$, and $a^{ijk}_3$ of octets may (and will) differ from those of decuplets.

As before, we can associate the $a^{ijk}_1-2a^{ijk}_2$ term with the bare mass of quarks:
\begin{equation}\label{eq-23}
a^{ijk}_1-2a^{ijk}_2 = m_k - \frac{1}{2}\left( m_i+m_j \right),
\end{equation}
That is, with each increase in $n_k$, one adds one $k$ quark, and removes either one $i$ or $j$ quark, and therefore the mass accordingly increases by $m_k$, and decreases (on average) by the average of $m_i$ and $m_j$. We can also associate the mass difference  ($\Delta M$) between two members of a generalized mass group as being due to the difference in the bare masses of the $i$ and $j$ quarks (assuming electromagnetic interactions can be neglected):
\begin{equation}\label{eq-24}
\Delta M = \Delta n_i m_i + \Delta n_j m_j.
\end{equation}
Noting that within a generalized mass group we have
\begin{equation}\label{EQ-TEMP}
\Delta n_i = -\Delta n_j,
\end{equation}
with the help of Eq.~(\ref{eq-17}), Eq.~(\ref{eq-24}) can be rewritten as
\begin{equation}\label{eq-25}
\Delta M = \Delta I^{ij}_\mathrm{z} \left( m_i - m_j \right).
\end{equation}
In the original GMO formalism, $\Delta M$ is considered\linebreak negligible. This is justified, as ${m_\mathrm{u} - m_\mathrm{d} \approx 0}$.

This model allows us to come up with good descriptions of how badly broken symmetries are in a given baryon multiplet. The following ratio:\footnote{The minus sign before $\frac{\Delta M}{\Delta I^{ij}_\mathrm{z}}$ is added so that the ratio is positive when $m_j > m_i$.}
\begin{equation}\label{eq-26}
Q^{ijk}_\mathrm{abs} = -\frac{\Delta M}{\Delta I^{ij}_\mathrm{z}} = a^{ijk}_3 = -\left( m_i - m_j \right),
\end{equation}
representing the ``slope'' of mass groups in a mass vs. $I^{ij}_\mathrm{z}$ graph, is a good absolute descriptor of the quality of the symmetry of the $i$ and $j$ in a given baryon multiplet. However, a relative (scale-independent) descriptor would be a more objective assessment of the quality of the symmetry. A convenient way to build such a descriptor would be to take the ratio of $Q^{ijk}_\mathrm{abs}$ relative to the ``mass height'' of the multiplet. For decuplets, using Eq.~(\ref{eq-19}) , Eq.~(\ref{eq-23}), and Eq.~(\ref{eq-26}), yields the ratio
\begin{subequations}
\begin{align}
Q^{ijk}_\mathrm{rel} \equiv \frac{Q^{ijk}_\mathrm{abs}}{\Omega_{ijk}-\Delta_{ijk}} & = +\frac{1}{3}\frac{a^{ijk}_3}{\left( a_1^{ijk} - 2a_2^{ijk} \right)} \label{eq-27a}\\ 
{} & = -\frac{1}{3} \frac{\left( m_i - m_j \right)}{\left[ m_k-\frac{1}{2}\left( m_i+m_j \right) \right]}.\label{eq-27b}
\end{align}
\end{subequations}
For octets, using Eq.~(\ref{eq-19}), Eq.~(\ref{eq-23}), and Eq.~(\ref{eq-26}), yields the ratio
\begin{subequations}
\begin{align}
Q^{ijk}_\mathrm{rel} \equiv \frac{Q^{ijk}_\mathrm{abs}}{\Xi_{ijk}-N_{ijk}} & = +\frac{a_3^{ijk}}{ 2 a_1^{ijk} - a_2^{ijk} } \label{eq-28a}\\ 
{} & \approx +\frac{1}{2}\frac{a_3^{ijk}}{\left( a_1^{ijk}-2a_2^{ijk} \right)} \label{eq-28b}\\ 
{} & \approx -\frac{1}{2} \frac{\left( m_i - m_j \right)}{\left[ m_k-\frac{1}{2}\left( m_i+m_j \right) \right]} \label{eq-28c}
\end{align}
\end{subequations}
with the approximation being valid when
\begin{equation}
a^{ijk}_1 \gg a^{ijk}_2,
\end{equation} 
or alternatively, when
\begin{equation}\label{APPROX}
\Xi_{ijk}-\Sigma_{ijk} \approx \Sigma_{ijk} - N_{ijk} \gg \Sigma_{ijk} - \Lambda_{ijk}.
\end{equation}
In practice, this approximation will always be valid except in the case of $uds$ octet, where it will underestimate its height by roughly 50\%. One could also rescale $Q^{ijk}_\mathrm{rel}$ by a factor of $\frac{2}{3}$ for the octet so that it compares better to the $Q^{ijk}_\mathrm{rel}$ for the decuplet:
\begin{equation}\label{eq-29}
{\tilde Q}^{ijk}_\mathrm{rel} \equiv
\left\{
\begin{matrix}
~ \frac{2}{3}Q^{ijk}_\mathrm{rel} & \approx -\frac{1}{3}\frac{\left( m_i - m_j \right)}{\left[ m_k - \frac{1}{2}\left( m_i + m_j \right) \right]} & \left( \mathrm{octet} \right), \\ 
 \\ 
~ Q^{ijk}_\mathrm{rel} & = -\frac{1}{3}\frac{\left( m_i - m_j \right)}{\left[ m_k - \frac{1}{2}\left( m_i + m_j \right) \right]} & \left( \mathrm{decuplet} \right).
\end{matrix}
\right.
\end{equation}
This will allow us to obtain an objective criteria for when it is physically meaningful to speak of generalized mass groups in both octets and decuplets.

\begin{figure*}[!t]
 \centering
 \begin{subfigure}{0.475\textwidth}
 \centering
 \includegraphics[width=\textwidth]{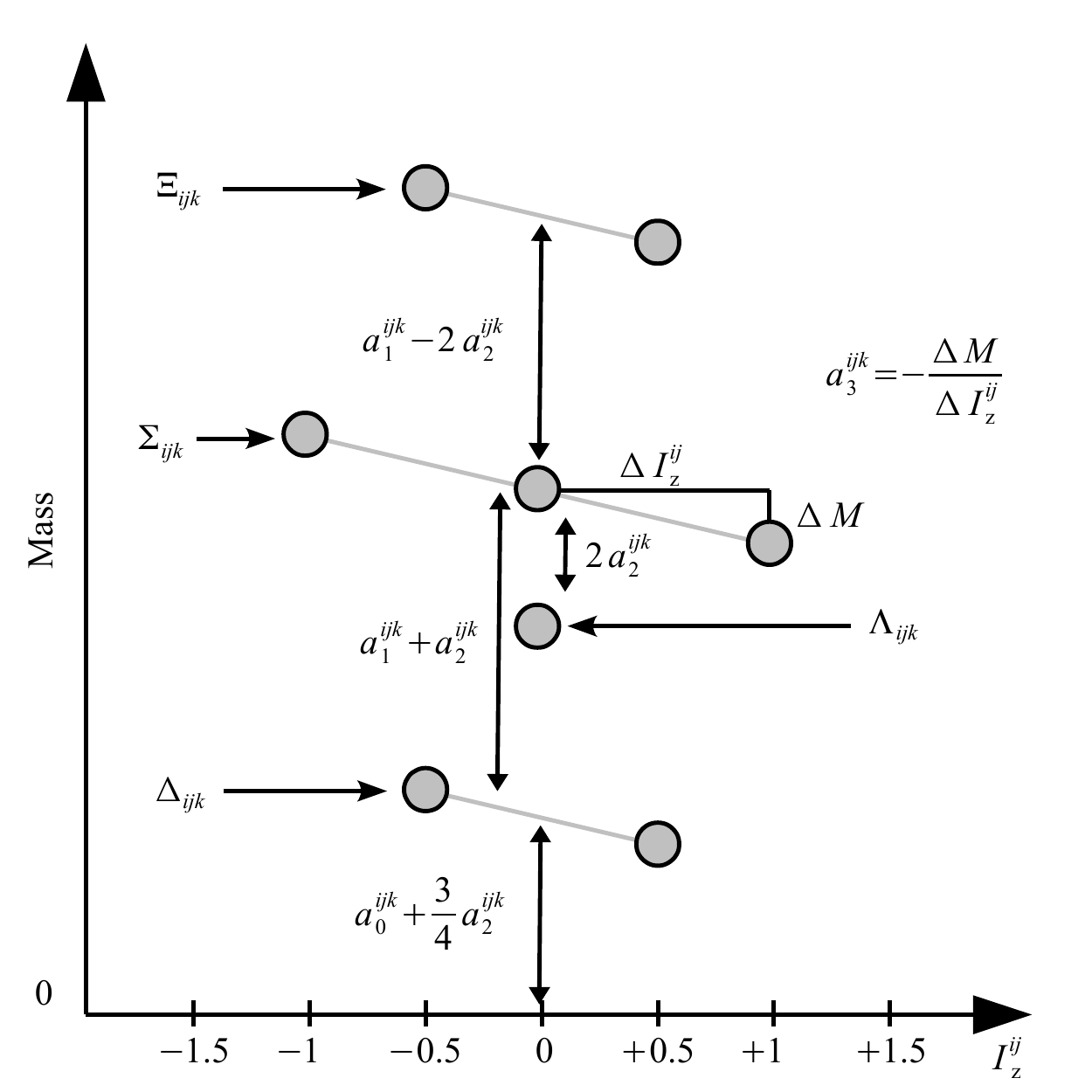}
 \caption{\label{Fig-GMOa}The generalized GMO parameters for octets.}
 \end{subfigure}
 \begin{subfigure}{0.475\textwidth}
 \centering
 \includegraphics[width=\textwidth]{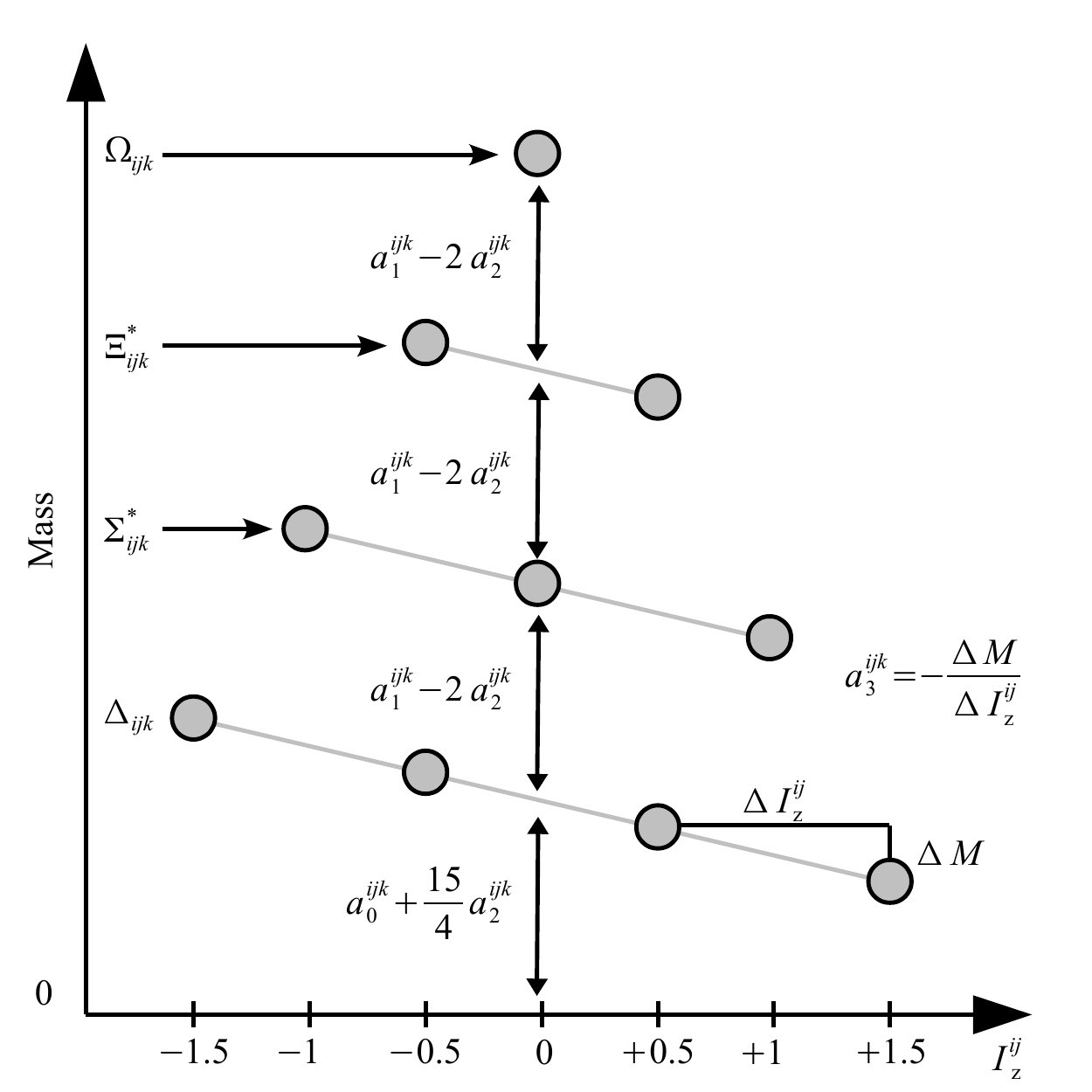}
 \caption{\label{Fig-GMOb}The generalized GMO parameters for decuplets.}
 \end{subfigure}
\caption{\label{Fig-GMO}The generalized GMO parameters and their relation to the masses of baryons.}
\end{figure*}

If, for instance, the $I^{ij}_\mathrm{z} = -\frac{3}{2}$ baryon of the $\Delta_{ijk}$ mass group is more massive than the $I^{ij}_\mathrm{z} = +1$ baryon of the $\Sigma^*_{ijk}$ mass group (see Fig.~\ref{Fig-GMOb}), it would be difficult to consider $\Delta_{ijk}$ and $\Sigma^*_{ijk}$ to be different mass groups. Therefore if the condition
\begin{equation}\label{eq-30}
\mathrm{max}(\Delta_{ijk}) < \mathrm{min}(\Sigma^*_{ijk})
\end{equation}
is satisfied, it will be meaningful to be speaking of mass groups. Expressing Eq.~(\ref{eq-30}) with the help of Eq.~(\ref{eq-25}) yields:
\begin{equation}\label{eq-31}
\Sigma^*_{ijk} - \Delta_{ijk} < -\frac{5}{2} \left( m_i - m_j \right)
\end{equation}
or, rewriting this in terms of the bare mass of quarks:
\begin{equation}\label{eq-32}
m_k - \frac{1}{2}\left(m_i-m_j\right) < -\frac{5}{2} \left( m_i - m_j \right).
\end{equation}
Substituting Eq.~(\ref{eq-32}) in Eq.~(\ref{eq-29}) yields the criterion
\begin{equation}\label{eq-33}
{\tilde Q}^{ijk}_\mathrm{rel} < \frac{2}{15} = 0.1\bar{3}.
\end{equation}
Other criteria, such as ${\mathrm{max}\left( \Sigma^*_{ijk} \right) < \mathrm{min}\left( \Xi^*_{ijk} \right)}$ are less restrictive, so we will not consider them.

A similar criteria can be devised in octets. If, for instance, the $I^{ij}_\mathrm{z} = -1$ baryon of the $\Sigma_{ijk}$ mass group is more massive than $I^{ij}_\mathrm{z} = +\frac{1}{2}$ baryon of the $\Xi_{ijk}$ mass group (see Fig.~\ref{Fig-GMOa}), it would be difficult to consider $\Xi_{ijk}$ and $\Sigma_{ijk}$ to be different mass groups. Therefore if the condition
\begin{equation}\label{eq-34}
\mathrm{max}(\Sigma_{ijk}) < \mathrm{min}(\Xi_{ijk})
\end{equation}
is satisfied, it will be meaningful to be speaking of mass groups. Expressing Eq.~(\ref{eq-34}) with the help of Eq.~(\ref{eq-25}) yields:
\begin{equation}\label{eq-35}
\Xi_{ijk} - \Sigma_{ijk} < -\frac{3}{2} \left( m_i - m_j \right)
\end{equation}
or, rewriting this in terms of the bare mass of quarks:
\begin{equation}\label{eq-36}
m_k - \frac{1}{2}\left(m_i-m_j\right) < -\frac{3}{2} \left( m_i - m_j \right).
\end{equation}
Substituting Eq.~(\ref{eq-36}) in Eq.~(\ref{eq-29}) yields the criterion
\begin{equation}\label{eq-37}
{\tilde Q}^{ijk}_\mathrm{rel} \lesssim \frac{2}{9} = 0.2\bar{2}.
\end{equation}
Considering instead ${\mathrm{max}\left( N_{ijk} \right) < \mathrm{min}\left( \Sigma_{ijk} \right)}$ would yield the same criterion, provided that the approximation in Eq.~(\ref{APPROX}) is valid. However, considering instead
\begin{equation}\label{eq-38}
\Lambda_{ijk} < \mathrm{min}\left( \Sigma_{ijk} \right)
\end{equation}
would, with the help of Eq.~(\ref{eq-25}) , yield
\begin{equation}\label{eq-39}
\Sigma_{ijk} - \Lambda_{ijk} < -\left( m_i - m_j \right).
\end{equation}
Using Eq.~(\ref{eq-split-2}) and Eq.~(\ref{eq-26}), we can express this as
\begin{equation}\label{eq-40}
2a_2^{ijk} > a_3^{ijk}.
\end{equation}
This is almost never satisfied in the octets. But since the $\Sigma_{ijk} - \Lambda_{ijk}$ splitting is only expected to be on the order of 50~MeV to 200~MeV, whereas the mass scales involved in the octets are on the order of anywhere from 1000~MeV to 12000~MeV, this is not a very meaningful criterion in the first place. Much more important is that $\Lambda_{ijk}$ is less massive than the $I^{ij}_\mathrm{z} = 0$ member of $\Sigma_{ijk}$. If this is satisfied, then $a_2^{ijk}$ will be positive.

We will see in the next section that the condition ${\tilde Q}_\mathrm{rel}^{ijk} < \frac{2}{15}$ is always satisfied in decuplets, and that both ${\tilde Q}_\mathrm{rel}^{ijk} \lesssim \frac{2}{9}$ and $a_2^{ijk}>0$ are always satisfied in octets, so long as the $i$, $j$, and $k$ quarks are chosen so that $k$ is the most massive of all.

\section{Determination of the generalized GMO parameters, $\tilde{Q}_\mathrm{rel}$, and the masses of all observable ${J^P = \frac{1}{2}^{+}}$ and ${J^P = \frac{3}{2}^{+}}$ baryons}
We analyzed each $ijk$ octet and decuplet individually, and obtained the best fit values for the generalized GMO parameters via least-square minimization using CurveFitter~4.5.8 \cite{CurveFitter}, with all data points given the same weight. We used all masses from the PDG particle listings \cite{PDG2012}, with the exception of the $\Xi_\mathrm{cc}^{+}$. The PDG lists the SELEX $\Xi^+_{cc}$ as having a mass of 3518.9~MeV, which would presumably have its isodoublet partner $\Xi_{cc}^{++}$ also lie in that mass range. However, it is given a 1-star rating for several reasons \cite{PDG2012}, and the mass value is so far off the predicted value (by\linebreak 160--200~MeV), both from our formalism and that of most others (see Table~XVII in \cite{Roberts2008} for a summary) that we do not consider it wise to include in our fits. All other baryons have either 4-star or 3-star ratings. The results are summarized in Table~\ref{tab-3} and Table~\ref{tab-4}.

In octets (see Table~\ref{tab-3}), the $a^{ijk}_0$ parameters do not have any obvious interpretation, but seem to be related to the mass of quarks in a subtle way. We see a pattern where increasing the mass of the $i$ and $j$ quarks increases the values of $a^{ijk}_0$, whereas increasing the mass of the $k$ quark decreases the values of $a^{ijk}_0$. We offer no explanation for this. Particularly reassuring is the fact that the $a^{ijk}_0$ values for the $ujk$ and $djk$ octets agree very closely to each other, as one would expect. The $a^{ijk}_1$ parameters on their own do not have any obvious interpretation, but the difference $a^{ijk}_1-2a^{ijk}_2$ should be related to the mass of quarks via Eq.~(\ref{eq-23}). And indeed we see that our results are in good agreement with what we can expect from the $\overline{\mathrm{MS}}$ masses of quarks, within 20\% in the $uds$ case, and within 10\% in the non-$uds$ cases. The $a^{ijk}_2$ parameters should be related to the $\Sigma_{ijk} - \Lambda_{ijk}$ splitting via Eq.~(\ref{eq-split-2}), and we have good agreement there, within 15~MeV in the $uds$ case, and within 1~MeV in the non-$uds$ cases. The $a_3^{ijk}$ parameters should be related to difference between the masses of the $i$ and $j$ quarks via Eq.~(\ref{eq-26}). However, they will also be sensitive to the electromagnetic interaction, which we neglected in our formalism, and which will affect the results by a few MeV. We should therefore expect no better than order-of-magnitude agreement in the $udk$ multiplets, and good agreement in the other cases. This is indeed the case. Lastly, the generalized GMO formalism accommodates the masses of octet baryons very well, with RMS errors under 10~MeV. However, in the case of non-$uds$ multiplets, the agreement could be purely fortuitous, as the masses of baryons containing two $c$ or $b$ quarks are unknown, and are required to put strong constraints on the generalized GMO parameters.

We could, however, predict the masses of the least massive of these baryons (as well as all others, see\linebreak Table~\ref{tab-PREDICT-O} and Fig.~\ref{fig-octets-all}) based on our fits. Using Eq.~(\ref{eq-19}) and the generalized GMO parameters from the $udc$ multiplet (see Table~\ref{tab-3}), the masses of $\Xi^{++}_\mathrm{cc}$ and $\Xi^{+}_\mathrm{cc}$ should be respectively 3717.46~MeV and 3717.62~MeV. Using the generalized GMO parameters from the $usc$ and $dsc$ multiplets, we instead get 3676.26~MeV and 3673.84~MeV respectively. Therefore, we expect the mass of $\Xi_\mathrm{cc}$ baryons to lie around $3697\pm30$~MeV, which would give us RMS errors on the order of 5~MeV. This is consistent with most predictions from other models (see Table~XVII in \cite{Roberts2008} for a summary). Other predictions from Table~\ref{tab-PREDICT-O} compare well with most other models as well (see the other tables in \cite{Roberts2008} for a summary). If, on the other hand, the SELEX results are confirmed, the predictive power of the generalized GMO formalism would be greatly diminished, and we could not claim RMS errors on the order of 5~MeV, but rather on the order of 40~MeV.

In decuplets, we have the relation (see Table~\ref{tab-2})
\begin{equation}\label{eq-41}
I^{ij} = \frac{3 - n_k}{2}
\end{equation}
which prevents us from obtaining $a^{ijk}_0$, $a^{ijk}_1$, and $a^{ijk}_2$ directly. However, we can rearrange Eq.~(\ref{eq-19}) by substituting Eq.~(\ref{eq-41}) into it to obtain:
\begin{equation}
\begin{split}\label{eq-42}
M \left(I^{ij},I^{ij}_\mathrm{z}, n_k \right) = & ~\left( a^{ijk}_0 + \frac{15}{4}a^{ijk}_2 \right) \\
{} & + \left( a^{ijk}_1 - 2a^{ijk}_2 \right) n_k\\
{} & - a^{ijk}_3 I^{ij}_\mathrm{z}
\end{split}
\end{equation}
The first term (constant) corresponds to the mass of the $\Delta_{ijk}$ group, the second term (linear in $n_k$) is the spacing of mass groups [Eq.~(\ref{eq-22})] and should be related to the mass of quarks via Eq.~(\ref{eq-23}), and the third (linear in $I^{ij}$) corresponds to the difference between the masses of the $i$ and $j$ quarks [Eq.~(\ref{eq-26})]. This is indeed the case (see Table~\ref{tab-4}). The remarks we made for $a^{ijk}_1-2a^{ijk}_2$ and $a_3^{ijk}$ in the octets apply to the decuplets as well. Again our results have good agreement with the expectations based on the $\overline{\mathrm{MS}}$ masses of quarks. The generalized GMO formalism again accommodates the masses of baryons well, although not as well as in the case of octets, this time with RMS errors typically on the order of 15~MeV. However, the uncertainties on baryon masses are also higher in decuplets, so the increase in RMS errors is to be expected.

To have an idea of how consistent the generalized GMO formalism is at predicting baryon masses in decuplets, we will take two test cases with unknown masses, but which are present in several decuplets: the $\Omega^{++}_\mathrm{ccc}$ and the $\Omega^{-}_\mathrm{bbb}$. Table~\ref{tab-PREDICT} and Fig.~\ref{fig-decuplets-all} summarize the predictions from the various multiplets using the generalized GMO parameters from Table~\ref{tab-4}. We see that there is remarkable agreement between multiplets for both of these baryons, despite the lack of strong constraints on the generalized GMO parameters, with the $\Omega^{++}_\mathrm{ccc}$ and the $\Omega^{-}_\mathrm{bbb}$ being predicted at $5062 \pm 40$~MeV and $15006 \pm 30$~MeV, respectively. We note that these are the \textit{least} consistent predictions in all decuplets, which span a scale from roughly 1200~MeV to 15000~MeV.

Lastly, we see in Table~\ref{tab-5} that ${{\tilde Q}_\mathrm{rel}^{ijk} < \frac{2}{15}}$ is always satisfied in decuplets, and that both ${{\tilde Q}_\mathrm{rel}^{ijk} \lesssim \frac{2}{9}}$ and ${a_2^{ijk}>0}$ are always satisfied in octets so long as the $i$, $j$, and $k$ quarks are chosen so that $k$ is the most massive of all.

\section{Conclusion}
We successfully generalized the familiar SU(3) framework of the $uds$ multiplets to any $ijk$ baryon multiplets via simple extensions of the existing concepts of isospin and mass groups. The generalized GMO formalism accommodates the masses of all observed baryons very well, and should allow for fairly accurate baryon mass predictions (with 50~MeV) for those not yet observed. The properties of baryons containing two $c$ or $b$ quarks will be a crucial test for the generalized isospin formalism.

It may be possible to use $a^{ijk}_0$ and the masses of the five observable quarks as free parameters, and do a fit on all 10 octets or 10 decuplets at once, rather than use $a^{ijk}_0$, $a^{ijk}_1$, $a^{ijk}_2$, and $a^{ijk}_3$ and do a fit on individual multiplets. This would reduce the number of free parameters from 40 to 15.

Regardless of the accuracy of the generalized GMO formalism, generalized isospin will at the very least allow for a flavor-independent, yet familiar, framework when working in the context of non-$udk$ baryon multiplets. It will also allow to distinguish between $\Lambda$-like baryons (part of a generalized isospin singlet, such as the $\Xi^+_\mathrm{c}$) and $\Sigma$-like baryons (part of a generalized isospin triplet, such as the $\Xi^{'+}_\mathrm{c}$) with a quantum number, instead of having to specify the nature of these baryons by comparing them to $udk$ analogs.

\begin{acknowledgments}
We acknowledge the financial support of Natural Sciences and Engineering Research Council of Canada (NSERC). The authors would also like to thank Marc Collette for his feedback and comments. This work is a continuation of \cite{Landry2013}, where the concepts of generalized isospin, generalized mass groups, and the generalized GMO formalism were first introduced by one of the authors.
\end{acknowledgments}

\bibliography{Isogen}

\begin{thebibliography}{18}%
\makeatletter
\providecommand \@ifxundefined [1]{%
 \@ifx{#1\undefined}
}%
\providecommand \@ifnum [1]{%
 \ifnum #1\expandafter \@firstoftwo
 \else \expandafter \@secondoftwo
 \fi
}%
\providecommand \@ifx [1]{%
 \ifx #1\expandafter \@firstoftwo
 \else \expandafter \@secondoftwo
 \fi
}%
\providecommand \natexlab [1]{#1}%
\providecommand \enquote  [1]{``#1''}%
\providecommand \bibnamefont  [1]{#1}%
\providecommand \bibfnamefont [1]{#1}%
\providecommand \citenamefont [1]{#1}%
\providecommand \href@noop [0]{\@secondoftwo}%
\providecommand \href [0]{\begingroup \@sanitize@url \@href}%
\providecommand \@href[1]{\@@startlink{#1}\@@href}%
\providecommand \@@href[1]{\endgroup#1\@@endlink}%
\providecommand \@sanitize@url [0]{\catcode `\\12\catcode `\$12\catcode
  `\&12\catcode `\#12\catcode `\^12\catcode `\_12\catcode `\%12\relax}%
\providecommand \@@startlink[1]{}%
\providecommand \@@endlink[0]{}%
\providecommand \url  [0]{\begingroup\@sanitize@url \@url }%
\providecommand \@url [1]{\endgroup\@href {#1}{\urlprefix }}%
\providecommand \urlprefix  [0]{URL }%
\providecommand \Eprint [0]{\href }%
\providecommand \doibase [0]{http://dx.doi.org/}%
\providecommand \selectlanguage [0]{\@gobble}%
\providecommand \bibinfo  [0]{\@secondoftwo}%
\providecommand \bibfield  [0]{\@secondoftwo}%
\providecommand \translation [1]{[#1]}%
\providecommand \BibitemOpen [0]{}%
\providecommand \bibitemStop [0]{}%
\providecommand \bibitemNoStop [0]{.\EOS\space}%
\providecommand \EOS [0]{\spacefactor3000\relax}%
\providecommand \BibitemShut  [1]{\csname bibitem#1\endcsname}%
\let\auto@bib@innerbib\@empty
\bibitem [{\citenamefont {Heisenberg}(1932{\natexlab{a}})}]{Heisenberg1932a}%
  \BibitemOpen
  \bibfield  {author} {\bibinfo {author} {\bibfnamefont {W.}~\bibnamefont
  {Heisenberg}},\ }\href {\doibase 10.1007/BF01342433} {\bibfield  {journal}
  {\bibinfo  {journal} {{Zeitschrift {f\"ur} Physik}}\ }\textbf {\bibinfo
  {volume} {77}},\ \bibinfo {pages} {1} (\bibinfo {year}
  {1932}{\natexlab{a}})}\BibitemShut {NoStop}%
\bibitem [{\citenamefont {Heisenberg}(1932{\natexlab{b}})}]{Heisenberg1932b}%
  \BibitemOpen
  \bibfield  {author} {\bibinfo {author} {\bibfnamefont {W.}~\bibnamefont
  {Heisenberg}},\ }\href {\doibase 10.1007/BF01337585} {\bibfield  {journal}
  {\bibinfo  {journal} {{Zeitschrift {f\"ur} Physik}}\ }\textbf {\bibinfo
  {volume} {78}},\ \bibinfo {pages} {156} (\bibinfo {year}
  {1932}{\natexlab{b}})}\BibitemShut {NoStop}%
\bibitem [{\citenamefont {Heisenberg}(1933)}]{Heisenberg1933}%
  \BibitemOpen
  \bibfield  {author} {\bibinfo {author} {\bibfnamefont {W.}~\bibnamefont
  {Heisenberg}},\ }\href {\doibase 10.1007/BF01335696} {\bibfield  {journal}
  {\bibinfo  {journal} {{Zeitschrift {f\"ur} Physik}}\ }\textbf {\bibinfo
  {volume} {80}},\ \bibinfo {pages} {587} (\bibinfo {year} {1933})}\BibitemShut
  {NoStop}%
\bibitem [{\citenamefont {Nakano}\ and\ \citenamefont
  {Nishijima}(1953)}]{Nakano1953}%
  \BibitemOpen
  \bibfield  {author} {\bibinfo {author} {\bibfnamefont {T.}~\bibnamefont
  {Nakano}}\ and\ \bibinfo {author} {\bibfnamefont {K.}~\bibnamefont
  {Nishijima}},\ }\href {\doibase 10.1143/PTP.10.581} {\bibfield  {journal}
  {\bibinfo  {journal} {{Progress of Theoretical Physics}}\ }\textbf {\bibinfo
  {volume} {10}},\ \bibinfo {pages} {581} (\bibinfo {year} {1953})}\BibitemShut
  {NoStop}%
\bibitem [{\citenamefont {Nishijima}(1955)}]{Nishijima1955}%
  \BibitemOpen
  \bibfield  {author} {\bibinfo {author} {\bibfnamefont {K.}~\bibnamefont
  {Nishijima}},\ }\href {\doibase 10.1143/PTP.13.285} {\bibfield  {journal}
  {\bibinfo  {journal} {{Progress of Theoretical Physics}}\ }\textbf {\bibinfo
  {volume} {13}},\ \bibinfo {pages} {285} (\bibinfo {year} {1955})}\BibitemShut
  {NoStop}%
\bibitem [{\citenamefont {Gell-Mann}(1956)}]{Gellmann1956}%
  \BibitemOpen
  \bibfield  {author} {\bibinfo {author} {\bibfnamefont {M.}~\bibnamefont
  {Gell-Mann}},\ }\href {\doibase 10.1007/BF02748000} {\bibfield  {journal}
  {\bibinfo  {journal} {{Il Nuovo Cimento Supplement}}\ }\textbf {\bibinfo
  {volume} {4}},\ \bibinfo {pages} {848} (\bibinfo {year} {1956})}\BibitemShut
  {NoStop}%
\bibitem [{\citenamefont {Gell-Mann}(1961)}]{Gellmann1961}%
  \BibitemOpen
  \bibfield  {author} {\bibinfo {author} {\bibfnamefont {M.}~\bibnamefont
  {Gell-Mann}},\ }\href@noop {} {\emph {\bibinfo {title} {{﻿The Eightfold
  Way: A Theory of Strong Interaction Symmetry}}}},\ \bibinfo {type} {Tech.
  Rep.}\ \bibinfo {number} {CTSL-20}\ (\bibinfo  {institution} {{California
  Institute of Technology, Synchrotron Laboratory}},\ \bibinfo {year}
  {1961})\BibitemShut {NoStop}%
\bibitem [{\citenamefont {Ne'emann}(1961)}]{Neemann1961}%
  \BibitemOpen
  \bibfield  {author} {\bibinfo {author} {\bibfnamefont {Y.}~\bibnamefont
  {Ne'emann}},\ }\href {\doibase 10.1016/0029-5582(61)90134-1} {\bibfield
  {journal} {\bibinfo  {journal} {{Nuclear Physics}}\ }\textbf {\bibinfo
  {volume} {26}},\ \bibinfo {pages} {222} (\bibinfo {year} {1961})}\BibitemShut
  {NoStop}%
\bibitem [{\citenamefont {Okubo}(1962{\natexlab{a}})}]{Okubo1962a}%
  \BibitemOpen
  \bibfield  {author} {\bibinfo {author} {\bibfnamefont {S.}~\bibnamefont
  {Okubo}},\ }\href {\doibase 10.1143/PTP.27.949} {\bibfield  {journal}
  {\bibinfo  {journal} {{Progress of Theoretical Physics}}\ }\textbf {\bibinfo
  {volume} {27}},\ \bibinfo {pages} {949} (\bibinfo {year}
  {1962}{\natexlab{a}})}\BibitemShut {NoStop}%
\bibitem [{\citenamefont {Okubo}(1962{\natexlab{b}})}]{Okubo1962b}%
  \BibitemOpen
  \bibfield  {author} {\bibinfo {author} {\bibfnamefont {S.}~\bibnamefont
  {Okubo}},\ }\href {\doibase 10.1143/PTP.28.24} {\bibfield  {journal}
  {\bibinfo  {journal} {{Progress of Theoretical Physics}}\ }\textbf {\bibinfo
  {volume} {28}},\ \bibinfo {pages} {24} (\bibinfo {year}
  {1962}{\natexlab{b}})}\BibitemShut {NoStop}%
\bibitem [{\citenamefont {Golberg}\ and\ \citenamefont
  {Leher-Ilamed}(1963)}]{Goldberg1963}%
  \BibitemOpen
  \bibfield  {author} {\bibinfo {author} {\bibfnamefont {H.}~\bibnamefont
  {Golberg}}\ and\ \bibinfo {author} {\bibfnamefont {Y.}~\bibnamefont
  {Leher-Ilamed}},\ }\href {\doibase 10.1063/1.1703982} {\bibfield  {journal}
  {\bibinfo  {journal} {{Journal of Mathematical Physics}}\ }\textbf {\bibinfo
  {volume} {4}},\ \bibinfo {pages} {501} (\bibinfo {year} {1963})}\BibitemShut
  {NoStop}%
\bibitem [{\citenamefont {Beringer}\ \emph {et~al.}(2012)\citenamefont
  {Beringer} \emph {et~al.}}]{PDG2012}%
  \BibitemOpen
  \bibfield  {author} {\bibinfo {author} {\bibfnamefont {J.}~\bibnamefont
  {Beringer}} \emph {et~al.},\ }\href {\doibase 10.1103/PhysRevD.86.010001}
  {\bibfield  {journal} {\bibinfo  {journal} {{Physical Review D}}\ }\textbf
  {\bibinfo {volume} {86}},\ \bibinfo {pages} {010001} (\bibinfo {year}
  {2012})},\ \bibinfo {note} {and 2013 partial update for the 2014 edition
  available at http://pdg.lbl.gov/}\BibitemShut {NoStop}%
\bibitem [{\citenamefont {Gell-Mann}(1962)}]{Gellmann1962}%
  \BibitemOpen
  \bibfield  {author} {\bibinfo {author} {\bibfnamefont {M.}~\bibnamefont
  {Gell-Mann}},\ }in\ \href@noop {} {\emph {\bibinfo {booktitle} {{Proceedings
  of the 1962 International Conference on High-Energy Physics at CERN}}}},\
  \bibinfo {editor} {edited by\ \bibinfo {editor} {\bibnamefont {{J.
  Prentki}}}}\ (\bibinfo  {publisher} {{CERN}},\ \bibinfo {year} {1962})\ p.\
  \bibinfo {pages} {805}\BibitemShut {NoStop}%
\bibitem [{\citenamefont {Barnes}\ \emph {et~al.}(1964)\citenamefont {Barnes}
  \emph {et~al.}}]{Barnes1964}%
  \BibitemOpen
  \bibfield  {author} {\bibinfo {author} {\bibfnamefont {V.~E.}\ \bibnamefont
  {Barnes}} \emph {et~al.},\ }\href {\doibase 10.1103/PhysRevLett.12.204}
  {\bibfield  {journal} {\bibinfo  {journal} {{Physical Review Letters}}\
  }\textbf {\bibinfo {volume} {12}},\ \bibinfo {pages} {204} (\bibinfo {year}
  {1964})}\BibitemShut {NoStop}%
\bibitem [{\citenamefont {Mahanthappa}\ and\ \citenamefont
  {Unger}(1977)}]{Mahanthappa1977}%
  \BibitemOpen
  \bibfield  {author} {\bibinfo {author} {\bibfnamefont {K.~T.}\ \bibnamefont
  {Mahanthappa}}\ and\ \bibinfo {author} {\bibfnamefont {D.~G.}\ \bibnamefont
  {Unger}},\ }\href {\doibase 10.1103/PhysRevD.16.3284} {\bibfield  {journal}
  {\bibinfo  {journal} {{Physical Review D}}\ }\textbf {\bibinfo {volume}
  {16}},\ \bibinfo {pages} {3284} (\bibinfo {year} {1977})}\BibitemShut
  {NoStop}%
\bibitem [{\citenamefont {Nylund}()}]{CurveFitter}%
  \BibitemOpen
  \bibfield  {author} {\bibinfo {author} {\bibfnamefont {L.}~\bibnamefont
  {Nylund}},\ }\href@noop {} {\enquote {\bibinfo {title} {{CurveFitter
  4.5.8}},}\ }\bibinfo {note} {(Institute of Mathematics and Statistics),
  available at http://www.math-solutions.org/curvefitter.html}\BibitemShut
  {NoStop}%
\bibitem [{\citenamefont {Roberts}\ and\ \citenamefont
  {Pervin}(2008)}]{Roberts2008}%
  \BibitemOpen
  \bibfield  {author} {\bibinfo {author} {\bibfnamefont {W.}~\bibnamefont
  {Roberts}}\ and\ \bibinfo {author} {\bibfnamefont {M.}~\bibnamefont
  {Pervin}},\ }\href {\doibase 10.1142/S0217751X08041219} {\bibfield  {journal}
  {\bibinfo  {journal} {{International Journal of Modern Physics A}}\ }\textbf
  {\bibinfo {volume} {23}},\ \bibinfo {pages} {2817} (\bibinfo {year}
  {2008})}\BibitemShut {NoStop}%
\bibitem [{\citenamefont {Landry}(2013)}]{Landry2013}%
  \BibitemOpen
  \bibfield  {author} {\bibinfo {author} {\bibfnamefont {G.}~\bibnamefont
  {Landry}},\ }\emph {\bibinfo {title} {{Sym{\'e}tries et nomenclature des
  baryons: proposition d'une nouvelle nomenclature}}},\ \href@noop {} {Master's
  thesis},\ \bibinfo  {school} {{Universit{\'e} de Moncton}} (\bibinfo {year}
  {2013})\BibitemShut {NoStop}%
\end{thebibliography}%

\newpage
\appendix
\section{\label{Appendix}Generalized isospin operators and related quantities}
Since generalized isospin is based on the mathematics of normal isospin, it will be completely isomorphic to isospin. In the $uds$ multiplets the ladder operators $\hat I_\pm$, $\hat V_\pm$, and $\hat U_\pm$, the z-component of isospin operator $\hat I_\mathrm{z}$, hypercharge operator $\hat Y$ and strangeness operator $\hat S$, acting on 
\begin{equation}
\vert u \rangle = \begin{pmatrix}1 \\ 0 \\ 0\end{pmatrix},~
\vert d \rangle = \begin{pmatrix}0 \\ 1 \\ 0\end{pmatrix},~
\vert s \rangle = \begin{pmatrix}0 \\ 0 \\ 1\end{pmatrix},
\end{equation}
are given in terms of the Gell-Mann matrices ($\lambda_i$) as
\begin{equation}
\renewcommand*{\arraystretch}{1.4}
\begin{matrix}
\hat I_{\pm} & = & \frac{1}{2}\left( \lambda_1 \pm \iota \lambda_2 \right), & \hat I_\mathrm{z} & = & \frac{1}{2}\lambda_3, \\ 
\hat V_{\pm} & = & \frac{1}{2}\left( \lambda_4 \pm \iota \lambda_5 \right), & \hat Y & = & \frac{1}{\sqrt{3}}\lambda_8, \\ 
\hat U_{\pm} & = & \frac{1}{2}\left( \lambda_6 \pm \iota \lambda_7 \right) & \hat S & = & + \frac{1}{\sqrt{3}}  \left( \lambda_8 - \frac{1}{\sqrt{3}} \mathbb{I} \right)
\end{matrix}
\end{equation}
where $\iota$ is the imaginary unit and $\mathbb{I}$ is the identity matrix. In the generalized isospin formalism, the equivalent operators, acting on
\begin{equation}
\vert i \rangle = \begin{pmatrix} 1 \\ 0 \\ 0 \end{pmatrix},~
\vert j \rangle = \begin{pmatrix} 0 \\ 1 \\ 0 \end{pmatrix}, ~
\vert k \rangle = \begin{pmatrix} 0 \\ 0 \\ 1 \end{pmatrix},
\end{equation}
would generalize to
\begin{equation}
\renewcommand*{\arraystretch}{1.4}
\begin{matrix}
\hat I^{ij}_{\pm} & = & \frac{1}{2}\left( \lambda_1 \pm \iota \lambda_2 \right), & \hat I^{ij}_\mathrm{z} & = & \frac{1}{2}\lambda_3, \\ 
\hat I^{ik}_{\pm} & = & \frac{1}{2}\left( \lambda_4 \pm \iota \lambda_5 \right), & \hat Y^{ij} & = & \frac{1}{\sqrt{3}}\lambda_8, \\ 
\hat I^{jk}_{\pm} & = & \frac{1}{2}\left( \lambda_6 \pm \iota \lambda_7 \right), & \hat K^{\phantom{ij}} & = & \pm \frac{1}{\sqrt{3}} \left( \lambda_8 - \frac{1}{\sqrt{3}} \mathbb{I} \right)
\end{matrix}
\end{equation}
with the sign of $\hat K$ being chosen so that it agrees with the charge of the quark of associated flavor. As such, all commutation relations and mathematical properties of isospin operators will also extend to the generalized isospin operators, and any result based on a $uds$ analysis should also hold to reasonable accuracy in a given $ijk$ multiplet, at least to the extent that the $ijk$ symmetries can be considered good (i.e., ${\tilde Q}^{ijk}_\mathrm{rel} \ll \frac{2}{15}$ in decuplets, or ${\tilde Q}^{ijk}_\mathrm{rel} \ll \frac{2}{9}$ in octets).

\newpage
\begin{table*}[!]
\caption{\label{tab-3}Generalized GMO parameters for octets\footnote{Parameters are given in $\mathrm{MeV}$ and are based on baryon masses from \protect\cite{PDG2012}. Plain values were determined using only the PDG baryon masses, while values in bold were estimated by completing the multiplet with the average baryon masses from Table~\protect\ref{tab-PREDICT-O}. Quarks masses used are the $\overline{\mathrm{MS}}$ masses from \protect\cite{PDG2012} and should only be used for order-of-magnitude considerations.}}
\renewcommand{\arraystretch}{1.4}
\begin{ruledtabular}
\begin{tabular}{cccccccccccc}
$ijk$ & $a_0^{ijk}$ & $a_1^{ijk}$ & $a_2^{ijk}$ & $a_3^{ijk}$ & $a_1^{ijk} - 2a_2^{ijk}$ & $m_k - \frac{1}{2}\left(m_i + m_j\right)$ & $2a^{ijk}_2$ & $\Sigma_{ijk} - \Lambda_{ijk}$ & $-\left(m_i - m_j\right)$ & RMS Error & Note \\ 
\hline
$uds$ & 911.33 & \phantom{0}200.83 & 44.60 & \phantom{000}4.05 & \phantom{0}111.63 & \phantom{00}91.45 & \phantom{0}89.2 & \phantom{0}76.96 & \phantom{000}2.5 & 6.95 & --- \\ 
$udc$ & 876.26 & 1431.08 & 83.54 & \phantom{000}0.16 & 1264.00 & 1271.45 & 167.08 & 166.44 & \phantom{000}2.5 & 0.49 & \footnotemark[2] \\ 
$udb$ & 866.17 & 4777.48 & 97.00 & \phantom{000}1.94 & 4583.48 & 4176.45 & 194.00 & --- & \phantom{000}2.5 & 0.23 & --- \\ 
$usc$ & 1211.94 & 1269.25 & 53.56 & \phantom{0}121.59 & 1162.12 & 1226.35 & 107.12 & 107.80 & \phantom{00}92.7 & 1.30 & --- \\ 
$usb$ & 1194.91 & 4608.41 & 76.27 & \phantom{0}125.49 & 4455.87 & 4131.35 & 152.54 & --- & \phantom{00}92.7 & 1.23 & \footnotemark[3] \\ 
$ucb$ & \textbf{3060.85} & \textbf{3959.33} & \textbf{19.43} & \textbf{1242.88} & \textbf{3920.47} & 3541.35 & \textbf{\phantom{0}38.86} & --- & 1272.3 & --- & \footnotemark[4] \\ 
$dsc$ & 1220.31 & 1263.67 & 52.37 & \phantom{0}121.44 & 1158.93 & 1225.1\phantom{0} & 104.74 & 107.02 & \phantom{00}90.2 & 1.46 & \footnotemark[2] \\ 
$dsb$ & 1202.52 & 4607.60 & 76.07 & \phantom{0}127.05 & 4457.19 & 4130.1\phantom{0} & 152.14 & --- & \phantom{00}90.2 & 0.99 & --- \\ 
$dcb$ & \textbf{3061.94} & \textbf{3965.65} & \textbf{17.08} & \textbf{1241.97} & \textbf{3931.49} & 3540.1\phantom{0} & \textbf{\phantom{0}34.16} & --- & 1270.2 & --- & \footnotemark[2] \footnotemark[4] \\ 
$scb$ & \textbf{3219.08} & \textbf{3890.75} & \textbf{35.74} & \textbf{1011.37} & \textbf{3819.27} & 3495\phantom{.00} & \textbf{\phantom{0}71.48} & --- & 1180.0 & --- & \footnotemark[4] \\ 
\end{tabular}
\footnotetext[2]{Excluding the $\Xi^+_{cc}$.}
\footnotetext[3]{$\Xi^0_\mathrm{b}$ mass estimated from the $\Xi^-_\mathrm{b}$ mass and $\Xi^0-\Xi^-$ mass splitting.}
\footnotetext[4]{Not enough data for a direct fit. Parameters were fitted using both PDG and Table~\protect\ref{tab-PREDICT-O} masses. Fit could accommodate any given mass.}
\end{ruledtabular}
\end{table*}

\begin{table*}[!]
\caption{\label{tab-4}Generalized GMO parameters for decuplets\footnote{Parameters are given in $\mathrm{MeV}$ and are based on baryon masses from \protect\cite{PDG2012}. Plain values were determined using only the PDG baryon masses, while values in bold were estimated by completing the multiplet with the average baryon masses from Table~\protect\ref{tab-PREDICT}. Quarks masses used are the $\overline{\mathrm{MS}}$ masses from \protect\cite{PDG2012} and should only be used for order-of-magnitude considerations.}}
\renewcommand{\arraystretch}{1.4}
\begin{ruledtabular}
\begin{tabular}{ccccccccc}
$ijk$ & $a_0^{ijk}+\frac{15}{4}a_2^{ijk}$ & $\Delta_{ijk}$ & $a_1^{ijk} - 2a_2^{ijk}$ & $m_k - \frac{1}{2}\left(m_i + m_j\right)$ & $a_3^{ijk}$ & $-\left(m_i - m_j\right)$ & RMS Error & Note \\
\hline
$uds$ & 1233.73 & 1232.00 & 148.37 & \phantom{00}91.45 & \phantom{000}0.80 & \phantom{000}2.5 & \phantom{0}3.18 & --- \\ 
$udc$ & 1232.00 & 1232.00 & 1286.07 & 1271.45 & \phantom{00}$-0.13$ & \phantom{000}2.5 & \phantom{0}0.33 & --- \\ 
$udb$ & 1232.00 & 1232.00 & 4601.60 & 4176.45 & \phantom{000}0.43 & \phantom{000}2.5 & \phantom{0}0.73 & --- \\ 
$usc$ & 1454.76 & 1455.01 & 1188.47 & 1226.36 & \phantom{0}140.45 & \phantom{00}92.7 & 10.67 & --- \\ 
$usb$ & 1454.76 & 1455.01 & 4506.03 & 4131.35 & \phantom{0}143.98 & \phantom{00}92.7 & \phantom{0}9.51 & --- \\ 
$ucb$ & 3160.85 & --- & 3957.15 & 3541.35 & 1285.90 & 1272.3 & --- & \footnotemark[2] \\ 
$dsc$ & 1456.66 & 1456.66 & 1186.77 & 1225.1\phantom{0} & \phantom{0}140.24 & \phantom{00}90.2 & 11.13 & --- \\ 
$dsb$ & 1456.66 & 1456.66 & 4507.18 & 4130.1\phantom{o} & \phantom{0}143.89 & \phantom{00}90.2 & \phantom{0}9.86 & \footnotemark[3] \\
$dcb$ & 3162.2 & --- & 3959.70 & 3540.1\phantom{0} & 1286.80 & 1270.2 & --- & \footnotemark[2] \\ 
$scb$ & \textbf{3355.12} & --- & \textbf{3886.07} & 3495\phantom{.00} & \textbf{1137.02} & 1180.0 & \textbf{17.51} & \footnotemark[4] \\ 
\end{tabular}
\footnotetext[2]{Fit could accommodate any given mass.}
\footnotetext[3]{$\Xi^{*0}_\mathrm{b}$ mass estimated from the $\Xi^{*-}_\mathrm{b}$ mass and $\Xi^{*0}-\Xi^{*-}$ mass splitting.}
\footnotetext[4]{Not enough data for a direct fit. Parameters were fitted using both PDG and Table~\protect\ref{tab-PREDICT} masses.}
\end{ruledtabular}
\end{table*}

\newpage
\begin{figure*}
 \centering
 \begin{subfigure}[b]{0.295\textwidth}
  \centering
  \includegraphics[width=\textwidth]{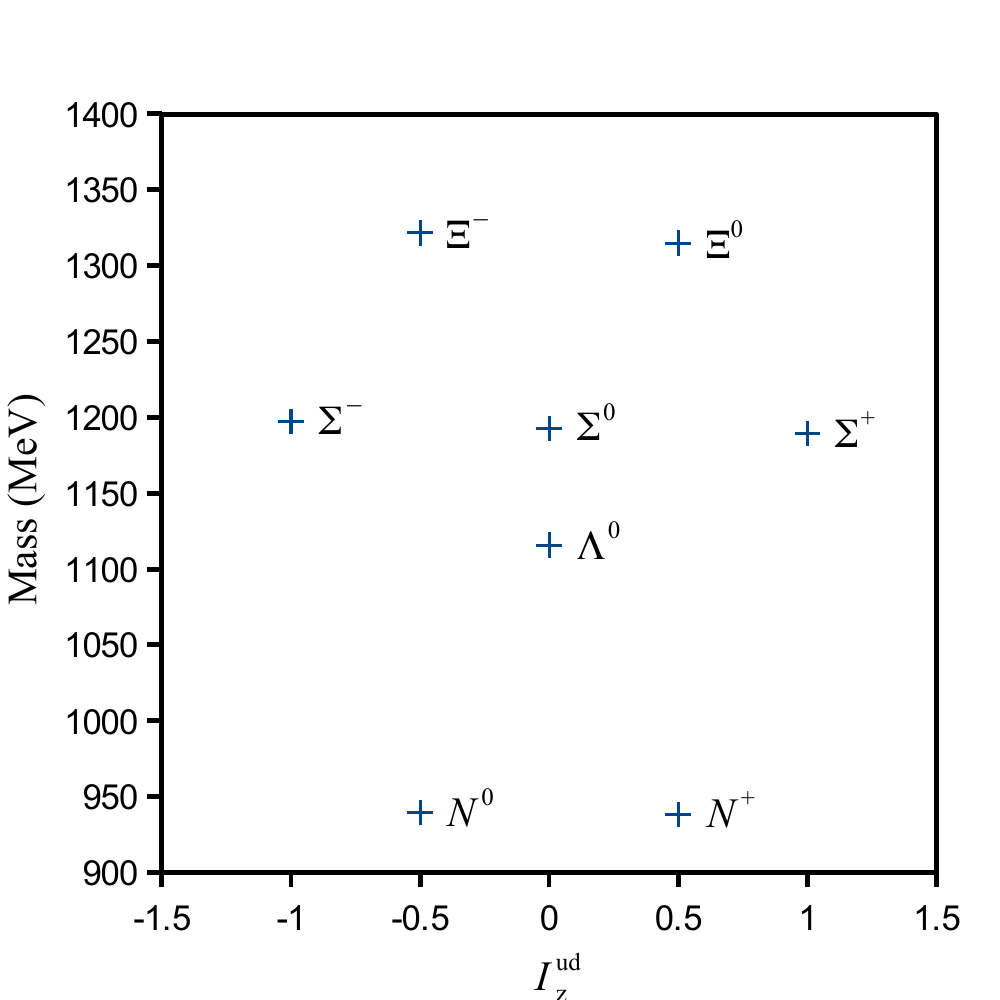}
  \caption{$uds$ octet}
  \label{fig-uds-octet}
 \end{subfigure}%
 \begin{subfigure}[b]{0.295\textwidth}
  \centering
  \includegraphics[width=\textwidth]{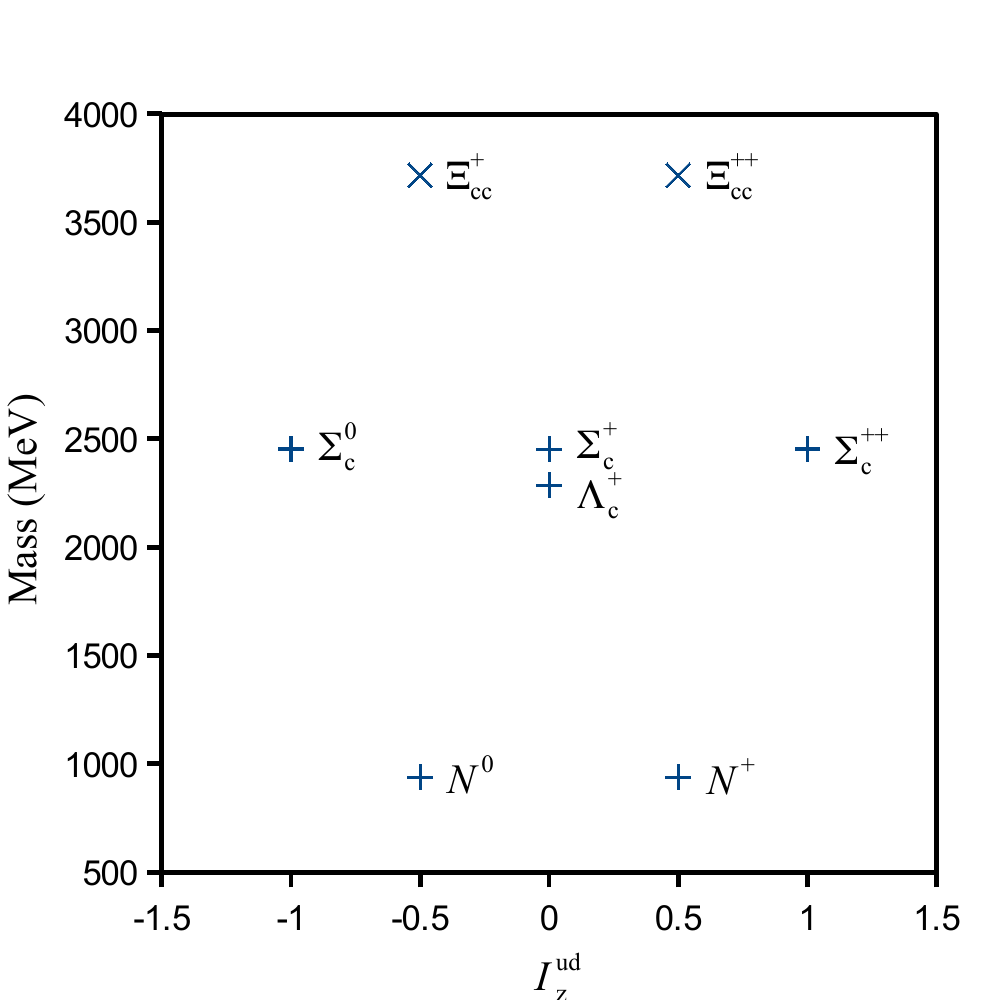}
  \caption{$udc$ octet}
  \label{fig-udc-octet}
 \end{subfigure}
 \begin{subfigure}[b]{0.295\textwidth}
  \centering
  \includegraphics[width=\textwidth]{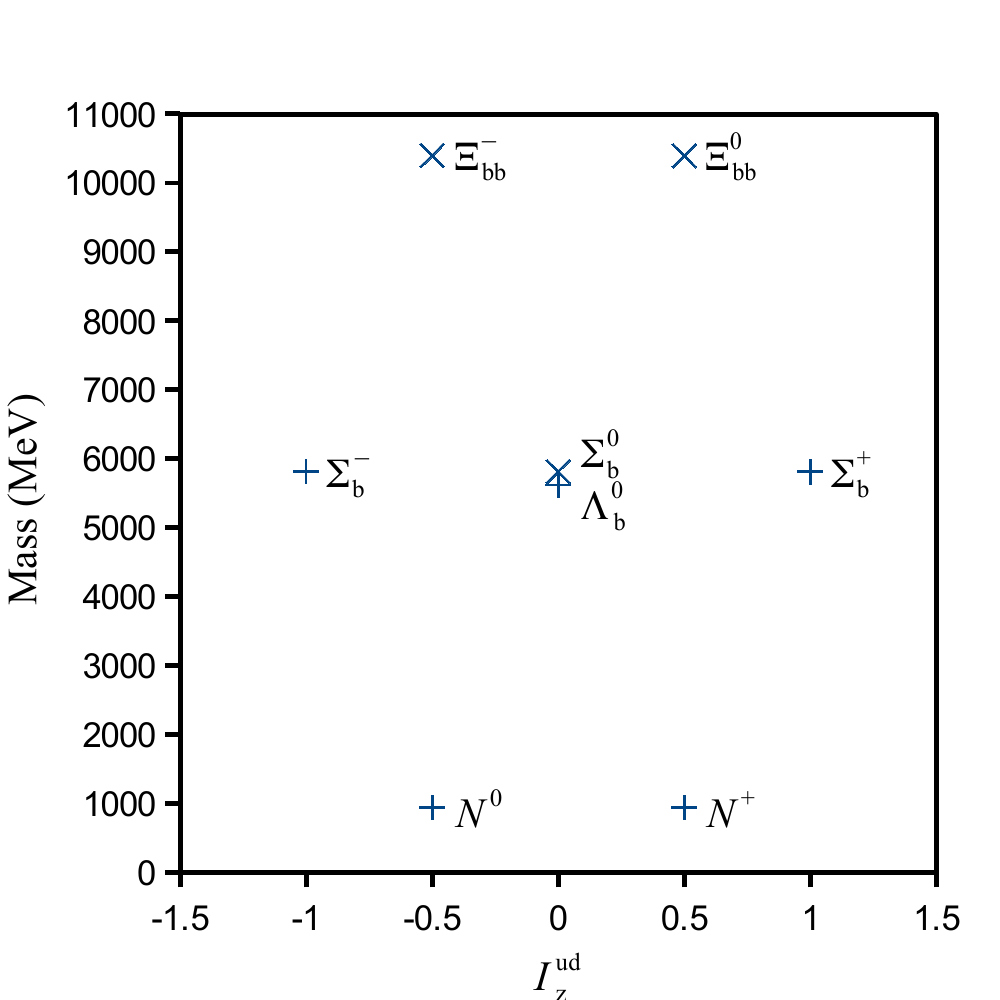}
  \caption{$udb$ octet}
  \label{fig-udb-octet}
 \end{subfigure}

 \begin{subfigure}[b]{0.295\textwidth}
  \centering
  \includegraphics[width=\textwidth]{usc-octet}
  \caption{$usc$ octet}
  \label{fig-usc-octet}
 \end{subfigure}%
 \begin{subfigure}[b]{0.295\textwidth}
  \centering
  \includegraphics[width=\textwidth]{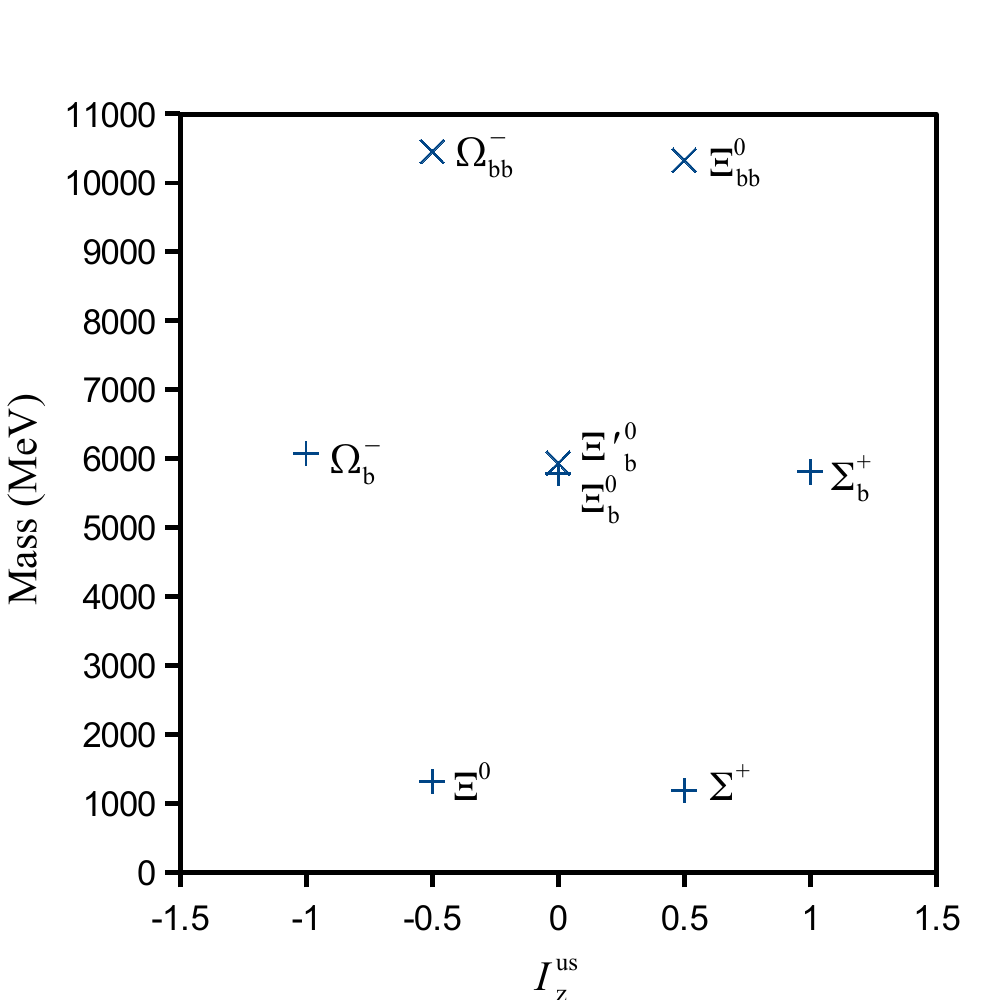}
  \caption{$usb$ octet}
  \label{fig-usb-octet}
 \end{subfigure}
 \begin{subfigure}[b]{0.295\textwidth}
  \centering
  \includegraphics[width=\textwidth]{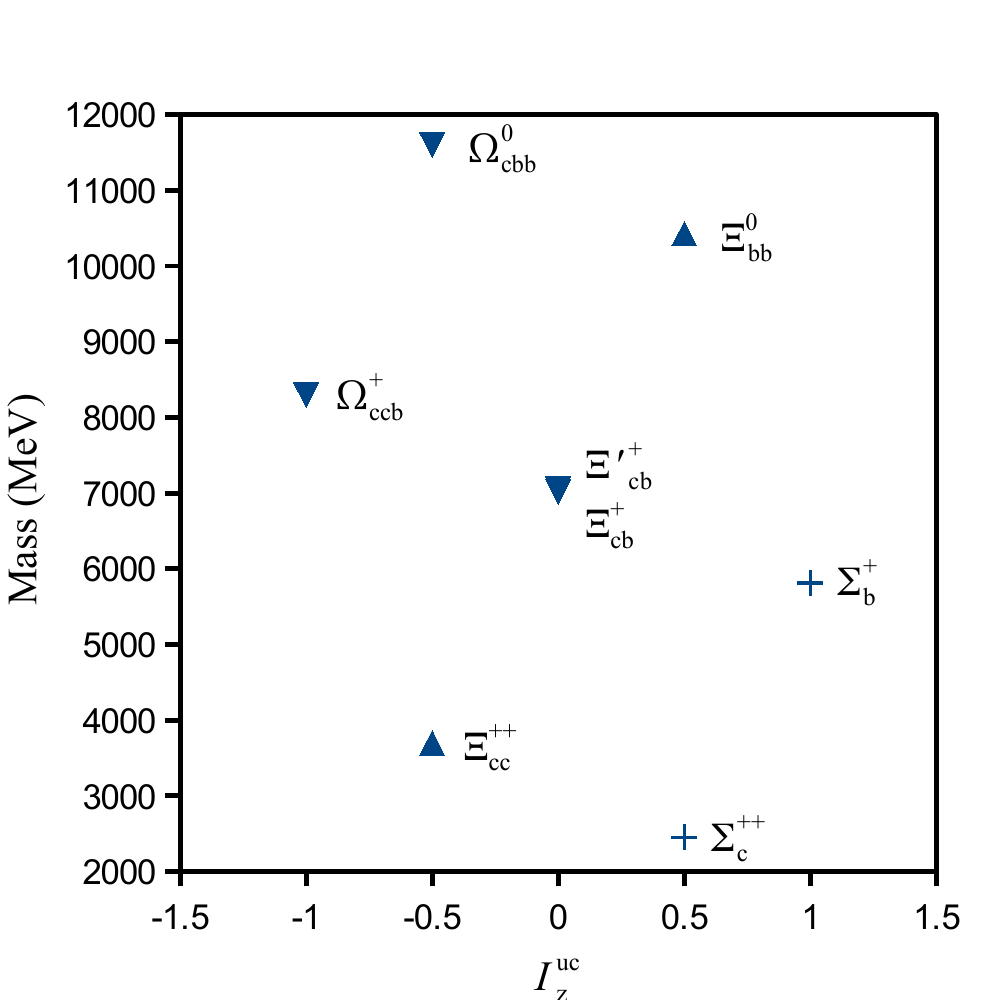}
  \caption{$ucb$ octet}
  \label{fig-ucb-octet}
 \end{subfigure}

 \begin{subfigure}[b]{0.295\textwidth}
  \centering
  \includegraphics[width=\textwidth]{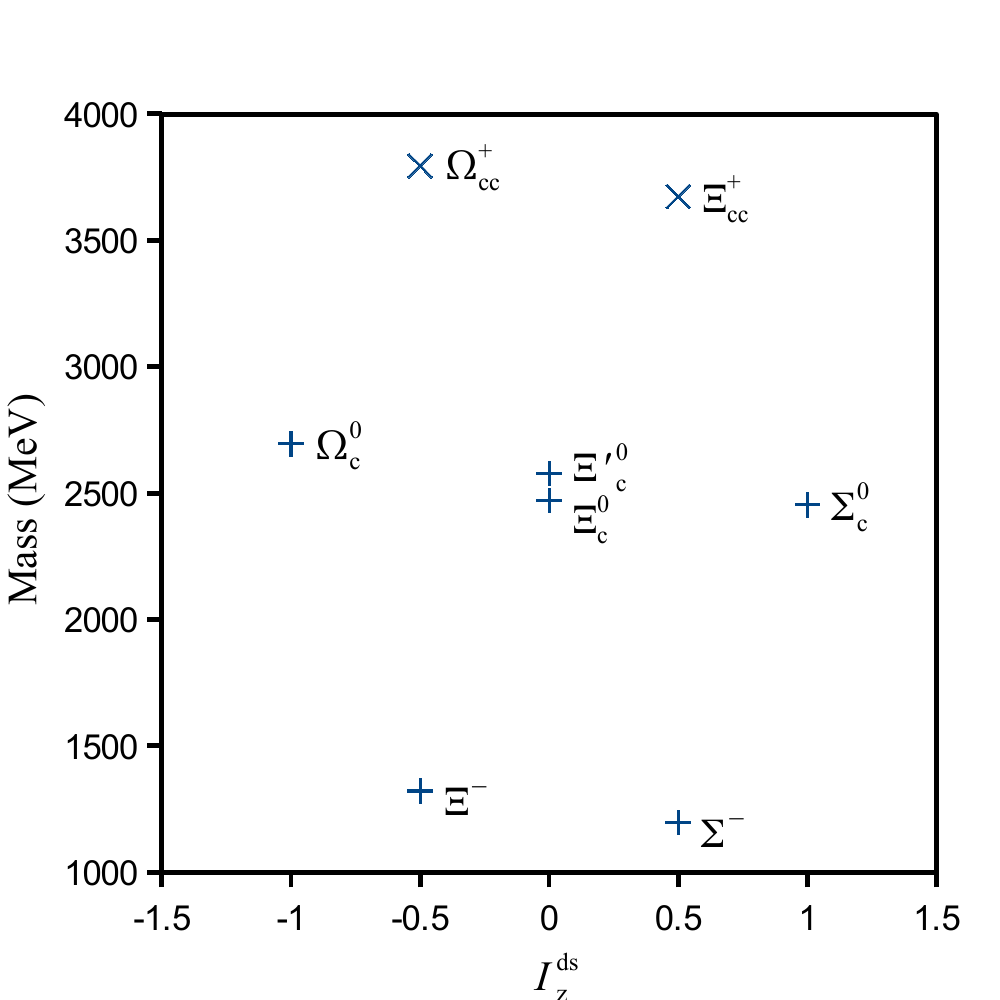}
  \caption{$dsc$ octet}
  \label{fig-dsc-octet}
 \end{subfigure}%
 \begin{subfigure}[b]{0.295\textwidth}
  \centering
  \includegraphics[width=\textwidth]{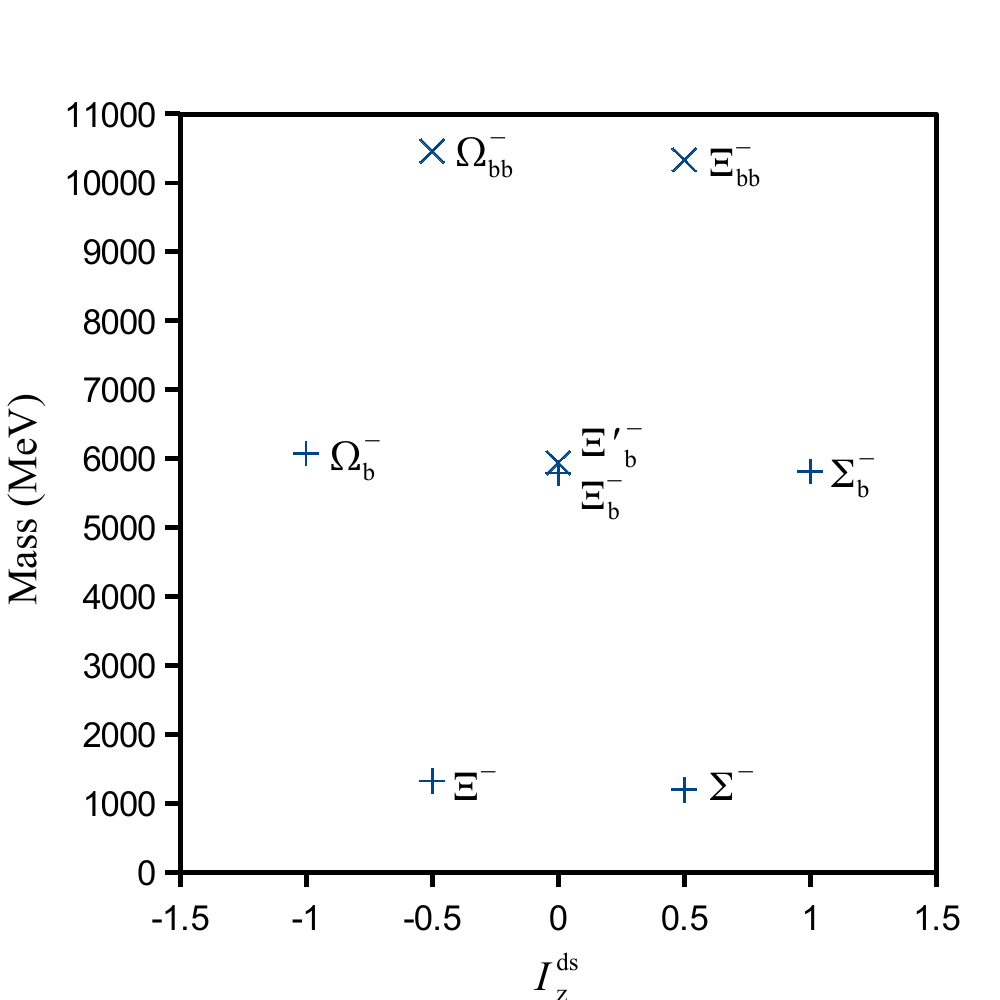}
  \caption{$dsb$ octet}
  \label{fig-dsb-octet}
 \end{subfigure}
 \begin{subfigure}[b]{0.295\textwidth}
  \centering
  \includegraphics[width=\textwidth]{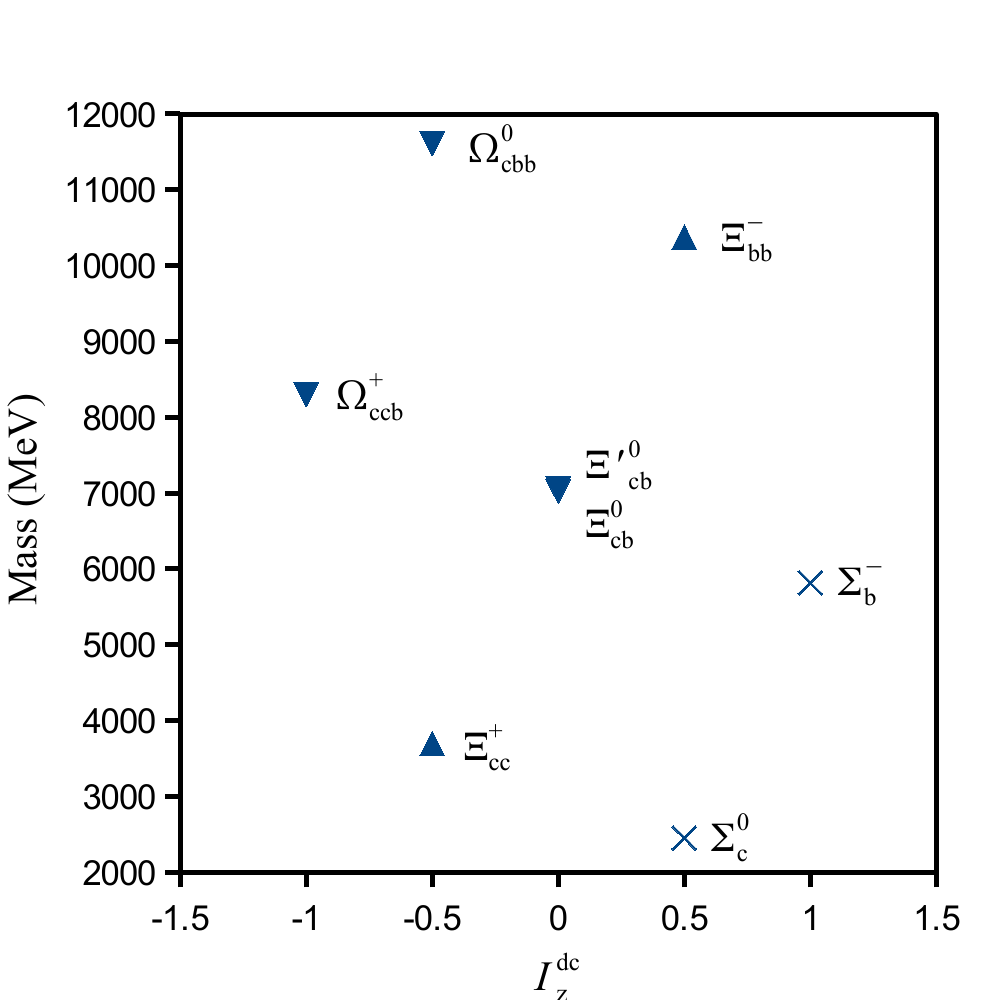}
  \caption{$dcb$ octet}
  \label{fig-dcb-octet}
 \end{subfigure}

 \begin{subfigure}[b]{0.295\textwidth}
  \centering
  \includegraphics[width=\textwidth]{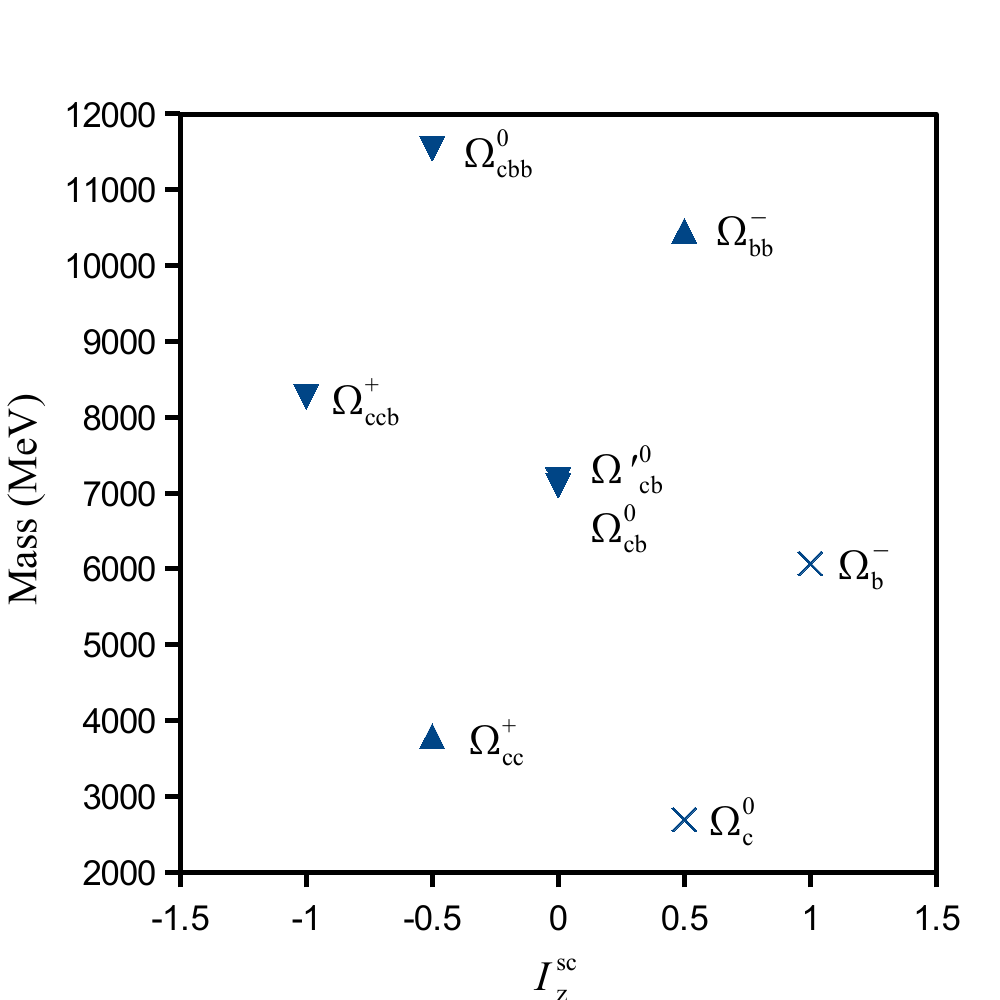}
  \caption{$scb$ octet}
  \label{fig-scb-octet}
  \end{subfigure}
\caption{\label{fig-octets-all}The $ijk$ octets. Baryon masses taken from \protect\cite{PDG2012} are marked +, while those taken from Table~\protect\ref{tab-PREDICT-O} are marked $\blacktriangle$. Mass estimates using Eq.~(\protect\ref{eq-19}) and direct parameter values from Table~\protect\ref{tab-3} are marked $\times$, while those using estimated values are marked $\blacktriangledown$.}
\end{figure*}

\newpage
\begin{figure*}
 \centering
 \begin{subfigure}[b]{0.295\textwidth}
  \centering
  \includegraphics[width=\textwidth]{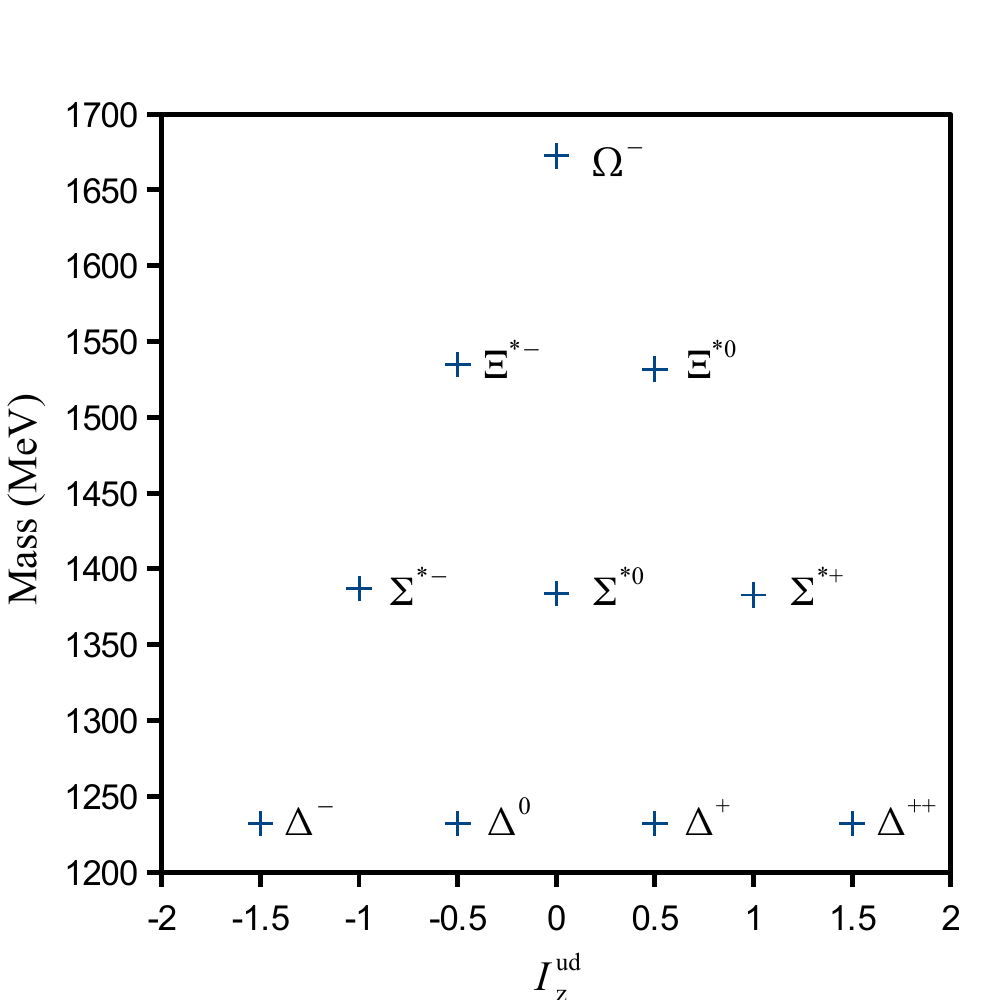}
  \caption{$uds$ decuplet}
  \label{fig-uds-decuplet}
  \end{subfigure}%
 \begin{subfigure}[b]{0.295\textwidth}
 \centering
  \includegraphics[width=\textwidth]{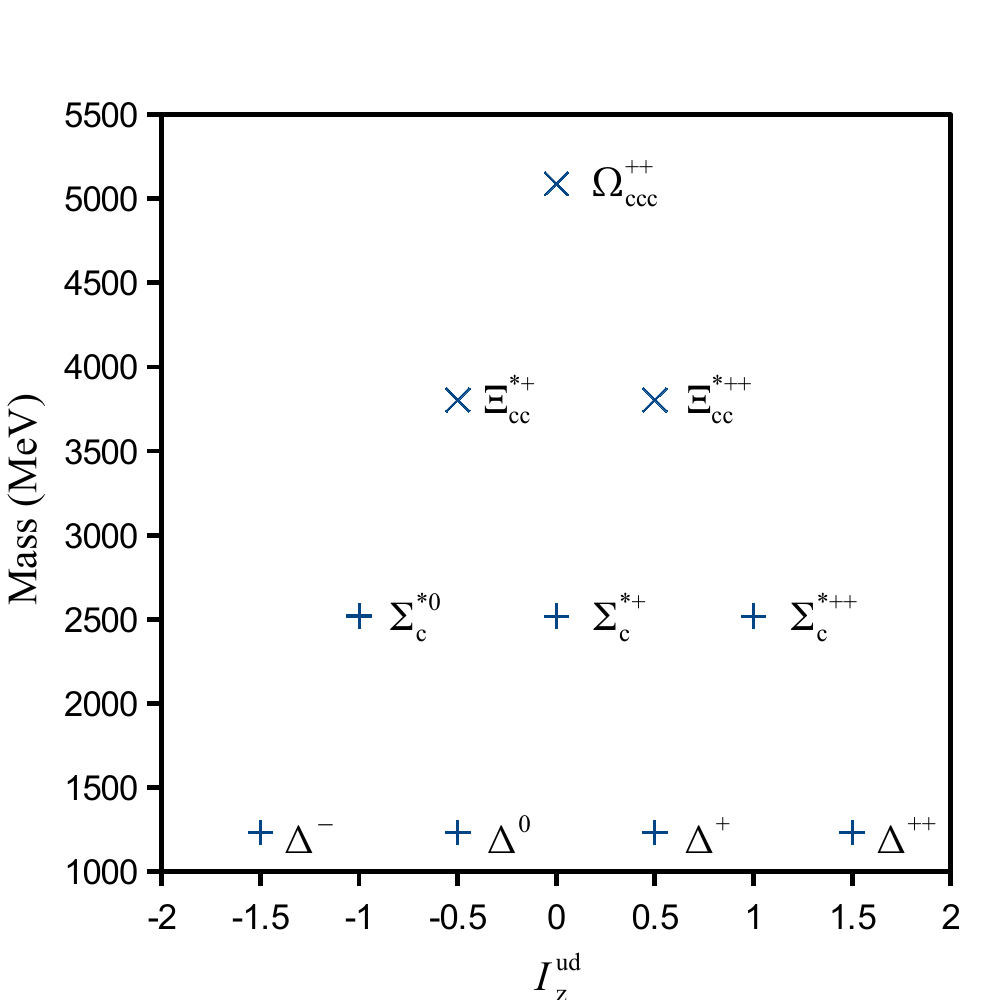}
  \caption{$udc$ decuplet}
  \label{fig-udc-decuplet}
  \end{subfigure}
 \begin{subfigure}[b]{0.295\textwidth}
 \centering
  \includegraphics[width=\textwidth]{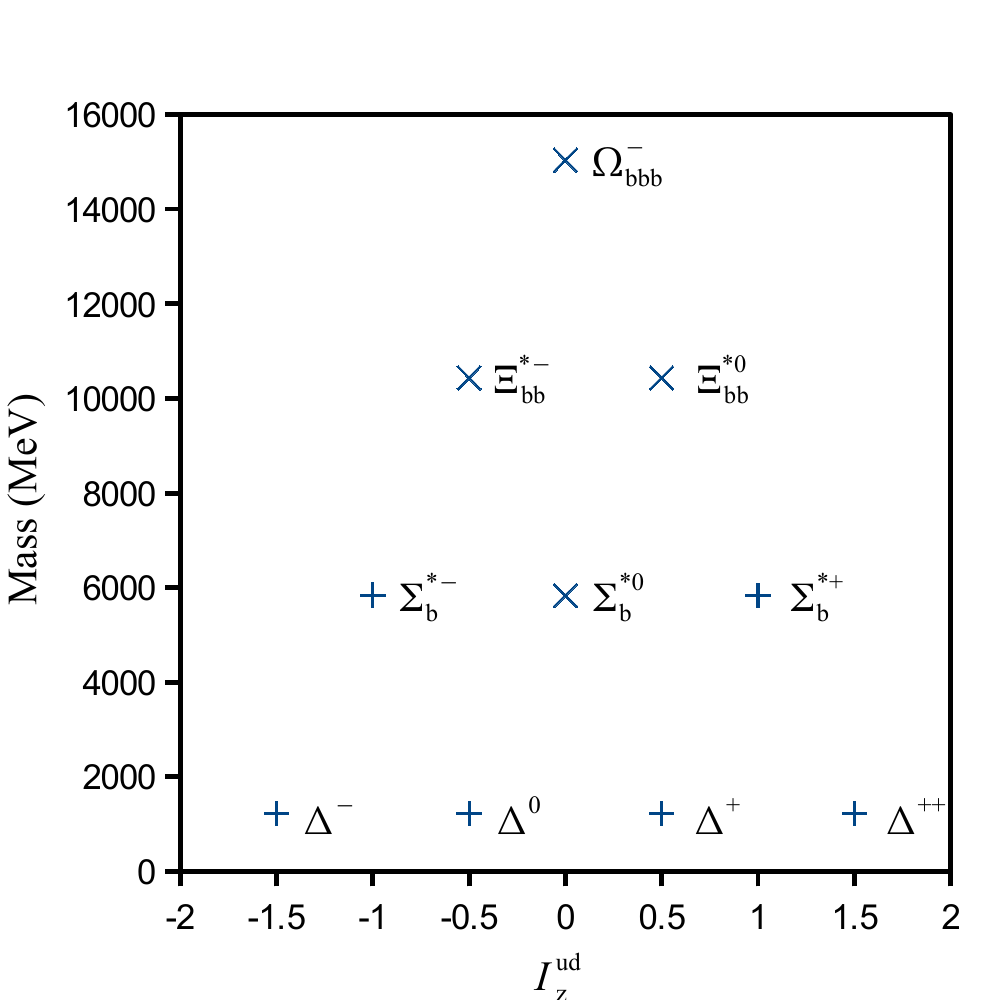}
  \caption{$udb$ decuplet}
  \label{fig-udb-decuplet}
 \end{subfigure}

 \begin{subfigure}[b]{0.295\textwidth}
  \centering
  \includegraphics[width=\textwidth]{usc-decuplet}
  \caption{$usc$ decuplet}
  \label{fig-usc-decuplet}
 \end{subfigure}%
 \begin{subfigure}[b]{0.295\textwidth}
  \centering
  \includegraphics[width=\textwidth]{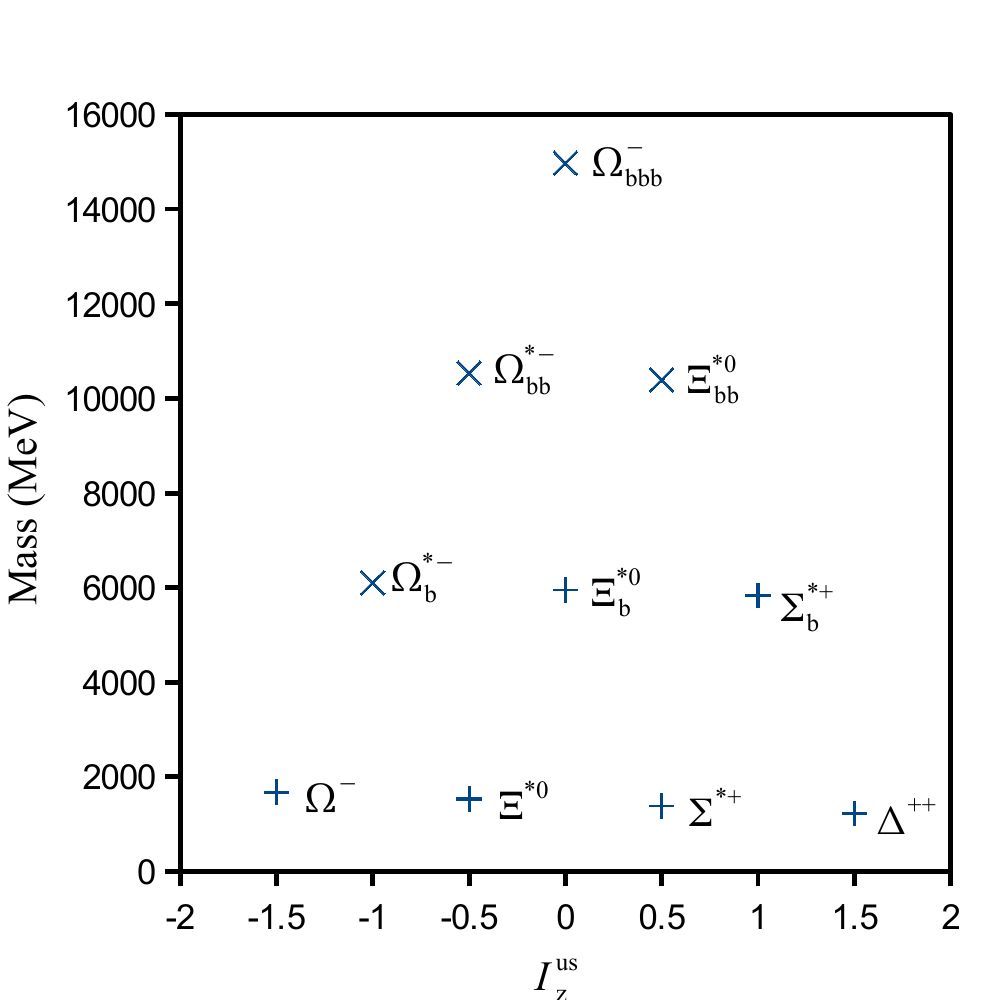}
  \caption{$usb$ decuplet}
  \label{fig-usb-decuplet}
 \end{subfigure}
 \begin{subfigure}[b]{0.295\textwidth}
  \centering
  \includegraphics[width=\textwidth]{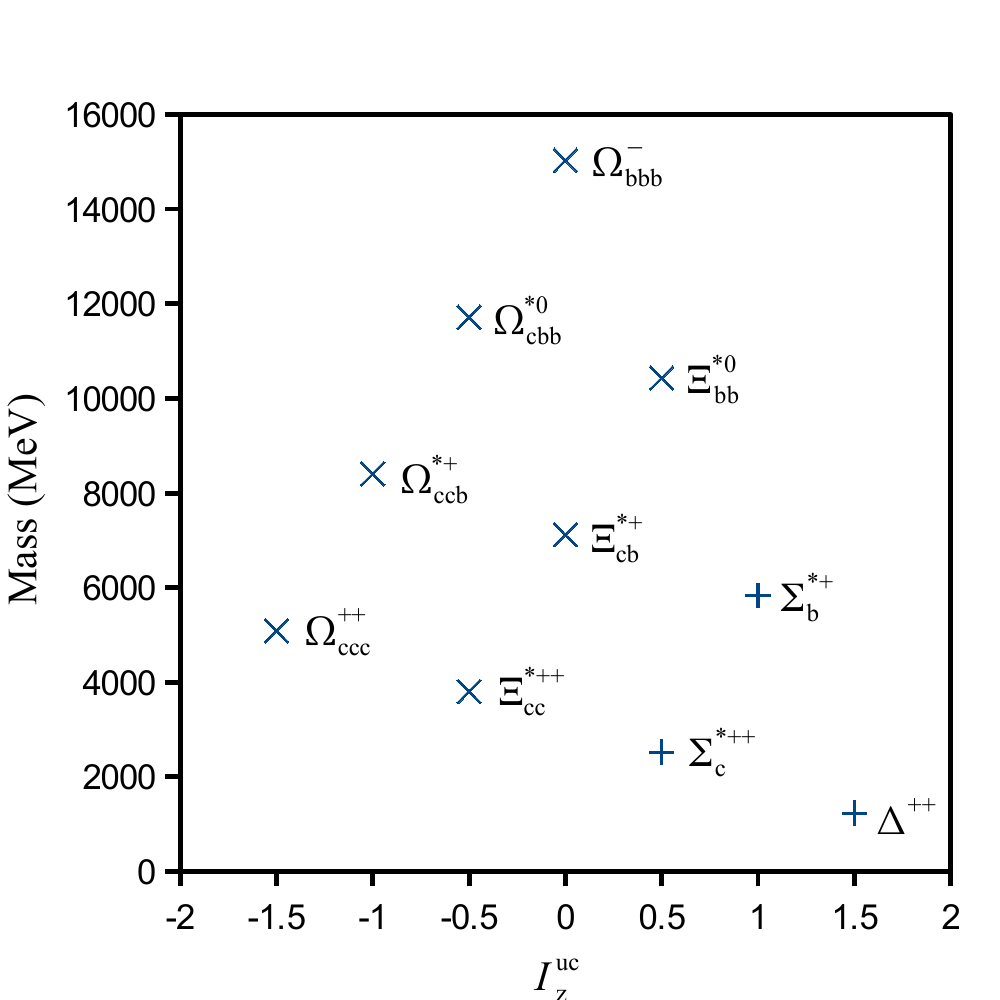}
  \caption{$ucb$ decuplet}
  \label{fig-ucb-decuplet}
 \end{subfigure}

 \begin{subfigure}[b]{0.295\textwidth}
  \centering
  \includegraphics[width=\textwidth]{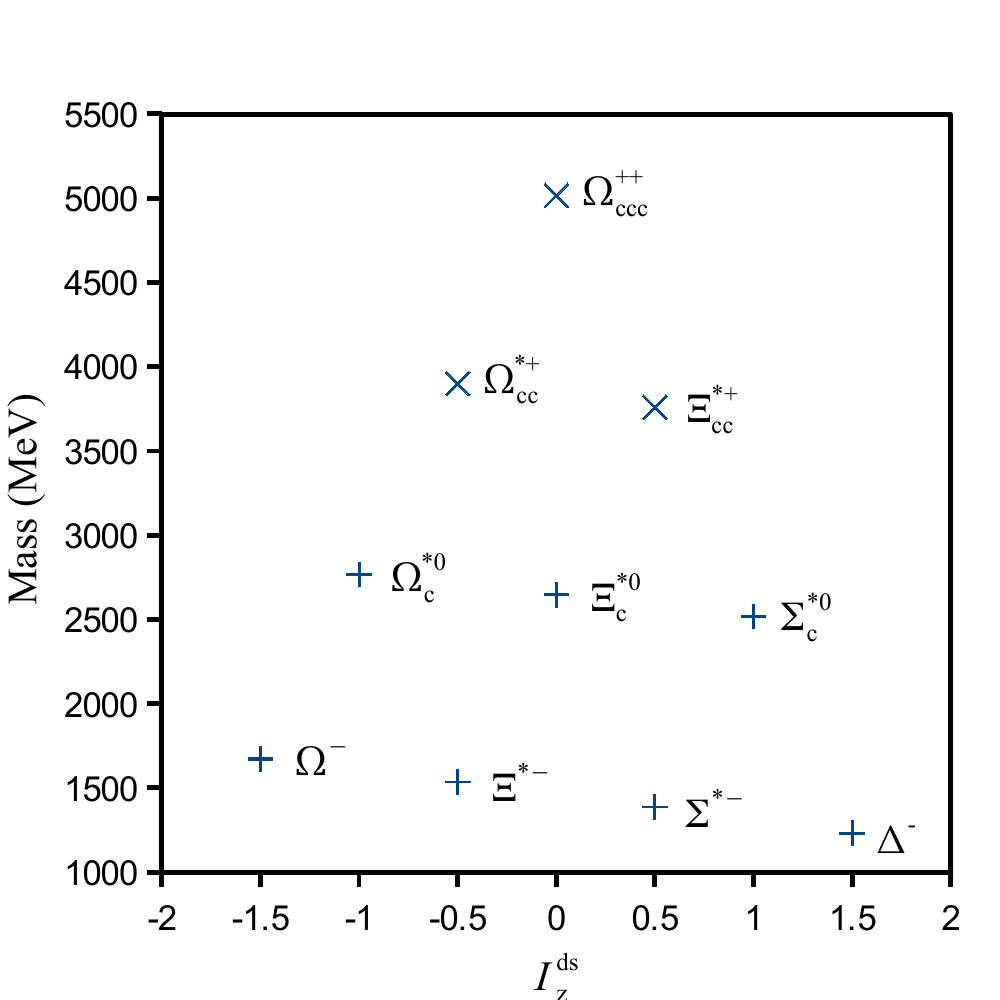}
  \caption{$dsc$ decuplet}
  \label{fig-dsc-decuplet}
 \end{subfigure}%
 \begin{subfigure}[b]{0.295\textwidth}
  \centering
  \includegraphics[width=\textwidth]{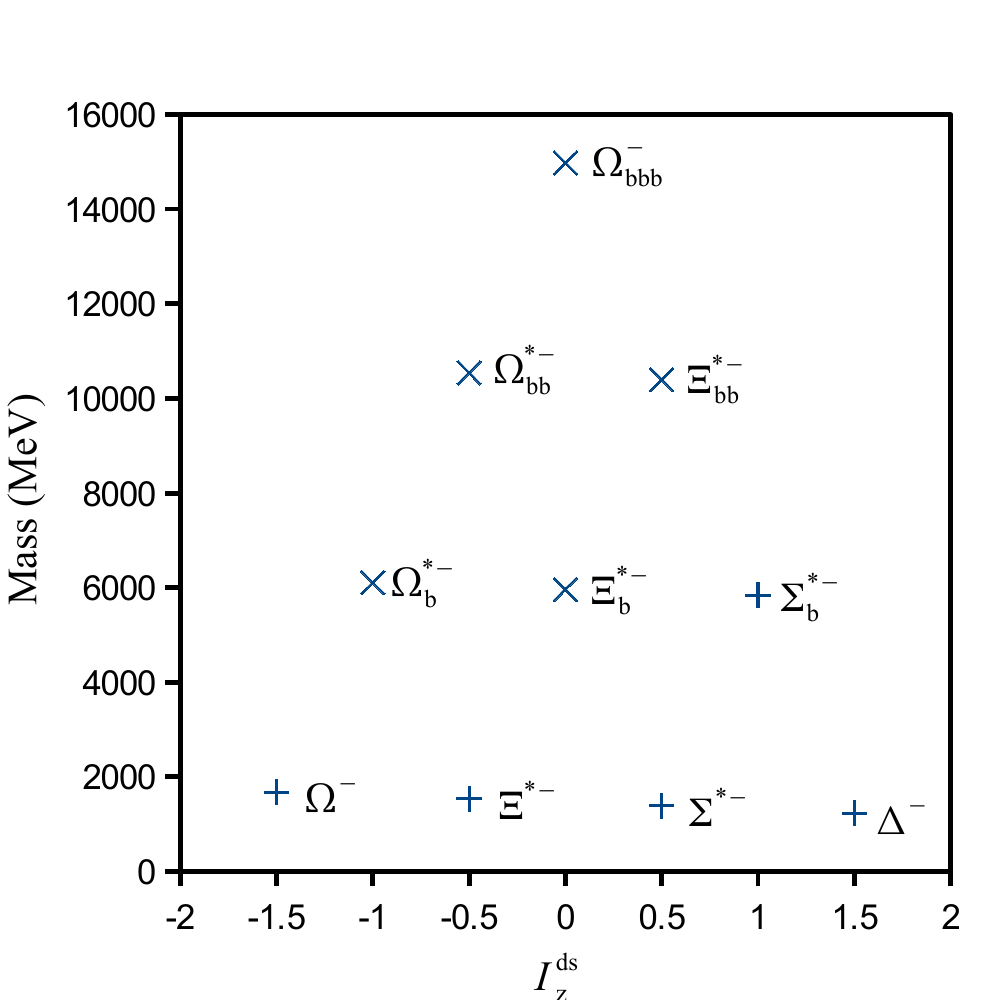}
  \caption{$dsb$ decuplet}
  \label{fig-dsb-decuplet}
 \end{subfigure}
 \begin{subfigure}[b]{0.295\textwidth}
  \centering
  \includegraphics[width=\textwidth]{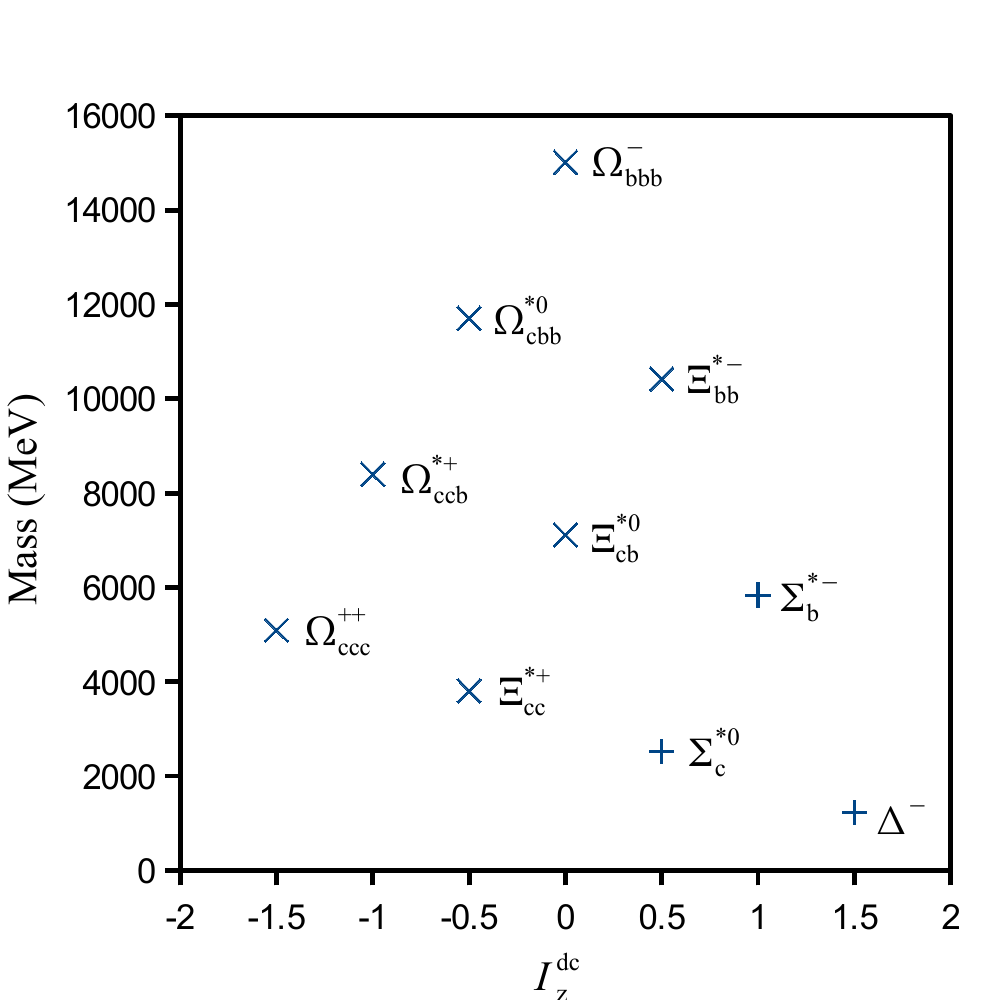}
  \caption{$dcb$ decuplet}
  \label{fig-dcb-decuplet}
 \end{subfigure}

 \begin{subfigure}[b]{0.295\textwidth}
  \centering
  \includegraphics[width=\textwidth]{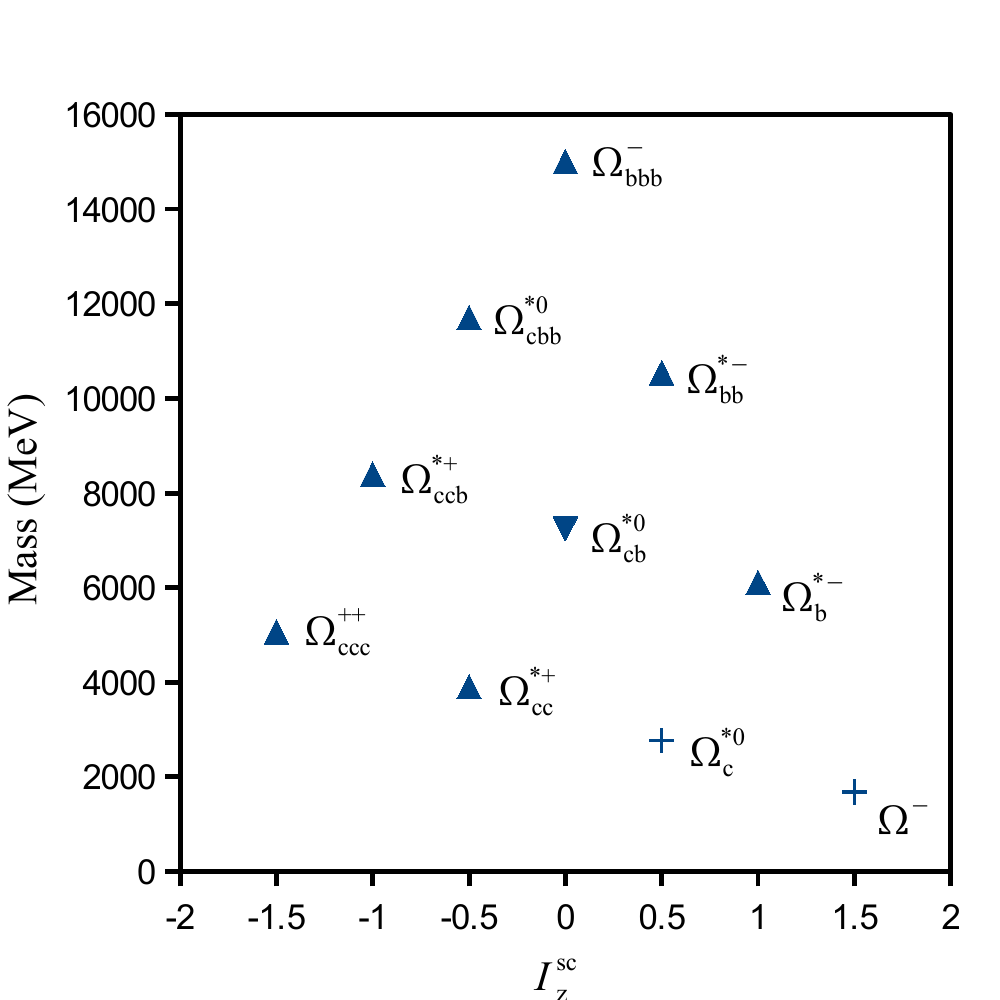}
  \caption{$scb$ decuplet}
  \label{fig-scb-decuplet}
 \end{subfigure}
\caption{\label{fig-decuplets-all}The $ijk$ decuplet. Baryon masses taken from \protect\cite{PDG2012} are marked +, while those taken from Table~\protect\ref{tab-PREDICT} are marked $\blacktriangle$. Mass estimates using Eq.~(\protect\ref{eq-19}) and direct parameter values from Table~\protect\ref{tab-4} are marked $\times$, while those using estimated values are marked $\blacktriangledown$.}
\end{figure*}

\newpage
\begin{table*}[h]
\caption{\label{tab-PREDICT-O}Predicted masses of missing octet baryons\footnote{Plain values were determined by fitting the generalized GMO parameters on the PDG baryon masses only. Values in bold could not be determined from PDG baryon masses only, and were instead determined by fitting the generalized GMO parameters on the PDG baryon masses as well as the average (plain) values from this table.}}
\renewcommand{\arraystretch}{1.4}
\begin{ruledtabular}
\begin{tabular}{cccccccccccc}
Multiplet & $\Omega^+_\mathrm{ccb}$ & Multiplet & $\Omega^0_\mathrm{cbb}$ & Multiplet & $\Xi^{++}_\mathrm{cc}$ & Multiplet & $\Xi^{+}_\mathrm{cc}$ & Multiplet & $\Xi^{0}_\mathrm{bb}$ & Multiplet & $\Xi^{-}_\mathrm{bb}$ \\
\cline{1-2} \cline{3-4} \cline{5-6} \cline{7-8} \cline{9-10} \cline{11-12}
$ucb$ & \textbf{8297.06} & $ucb$ & \textbf{11596.09} & $udc$ & 3717.46 & $udc$ & 3717.62 & $udb$ & 10395.91 & $udb$ & 10397.85 \\
$dcb$ & \textbf{8299.45} & $dcb$ & \textbf{11609.96} & $usc$ & 3676.25 & $dsc$ & 3673.84 & $usb$ & 10329.92 & $dsb$ & 10335.18 \\
$scb$ & \textbf{8273.75} & $scb$ & \textbf{11542.33} \\
\cline{1-2} \cline{3-4} \cline{5-6} \cline{7-8} \cline{9-10} \cline{11-12}
Average & \textbf{8290.09} & Average & \textbf{11582.79} & Average & 3696.86 & Average & 3695.73 & Average & 10362.92 & Average & 10376.52 \\
$\sigma$ & \textbf{\phantom{00}14.20} & $\sigma$ & \textbf{\phantom{000}35.72} & $\sigma$ & \phantom{00}29.14 & $\sigma$ & \phantom{00}30.96 & $\sigma$ & \phantom{000}46.66 & $\sigma$ & \phantom{000}30.17 \\ 
\hline
\hline
Multiplet & $\Omega^{+}_\mathrm{cc}$ & Multiplet & $\Omega^{-}_\mathrm{bb}$ & Multiplet & $\Sigma^0_\mathrm{b}$ & Multiplet & $\Xi^{'0}_\mathrm{b}$ & Multiplet & $\Xi^{'-}_\mathrm{b}$ & Multiplet & $\Xi^{+}_\mathrm{cb}$ \\
\cline{1-2} \cline{3-4} \cline{5-6} \cline{7-8} \cline{9-10} \cline{11-12}
$usc$ & 3797.85 & $usb$ & 10455.41 & $udb$ & 5813.40 & $usb$ & 5936.79 & $dsb$ & \phantom{0}5943.24 & $ucb$ & \textbf{\phantom{0}7015.32} \\
$dsc$ & 3795.28 & $dsb $ & 10463.23 & \\
\cline{1-2} \cline{3-4} \cline{5-6} \cline{7-8} \cline{9-10} \cline{11-12}
Average & 3796.57 & Average & 10458.82 & Average & 5813.40 & Average & 5936.79 & Average & \phantom{0}5943.24 & Average & \textbf{\phantom{0}7015.32} \\
$\sigma$ & \phantom{000}1.82 & $\sigma$ & \phantom{0000}4.82 & $\sigma$ & --- & $\sigma$ & --- & $\sigma$ & --- & $\sigma$ & --- \\
\hline
\hline
Multiplet & $\Xi^{'+}_\mathrm{cb}$ & Multiplet & $\Xi^{0}_\mathrm{cb}$ & Multiplet & $\Xi^{'0}_\mathrm{cb}$ & Multiplet & $\Omega^{0}_\mathrm{cb}$ & Multiplet & $ \Omega^{'0}_\mathrm{cb}$ \\
$ucb$ & \textbf{7054.18} & $dcb$ & \textbf{\phantom{0}7023.32} & $dcb$ & \textbf{7057.48} & $scb$ & \textbf{7100.90} & $scb$ & \textbf{\phantom{0}7172.38} \\
\cline{1-2} \cline{3-4} \cline{5-6} \cline{7-8} \cline{9-10} 
Average & \textbf{7054.18} & Average & \textbf{\phantom{0}7023.32} & Average & \textbf{7057.48} & Average & \textbf{7100.90} & Average & \textbf{\phantom{0}7172.38} \\
$\sigma$ & --- & $\sigma$ & --- & $\sigma$ & --- & $\sigma$ & --- & $\sigma$ & --- 
\end{tabular}
\end{ruledtabular}
\end{table*}

\begin{table*}[h]
\caption{\label{tab-PREDICT}Predicted masses of missing decuplet baryons\footnote{Plain values were determined by fitting the generalized GMO parameters on the PDG baryon masses only. Values in bold could not be determined from PDG baryon masses only, and were instead determined by fitting the generalized GMO parameters on the PDG baryon masses as well as the average (plain) values from this table.}}
\renewcommand{\arraystretch}{1.4}
\begin{ruledtabular}
\begin{tabular}{cccccccccccc}
Multiplet & $\Omega^{++}_\mathrm{ccc}$ & Multiplet & $\Omega^{-}_\mathrm{bbb}$ & Multiplet & $\Xi^{*++}_\mathrm{cc}$ & Multiplet & $\Xi^{*+}_\mathrm{cc}$ & Multiplet & $\Xi^{*0}_\mathrm{bb}$ & Multiplet & $\Xi^{*-}_\mathrm{bb}$ \\
\cline{1-2} \cline{3-4} \cline{5-6} \cline{7-8} \cline{9-10} \cline{11-12}
$udc$ & 5090.21 & $udb$ & 15036.80 & $udc$ & \phantom{0}3804.21 & $udc$ & 3804.08 & $udb$ & 10434.99 & $udb$ & 10435.45\\ 
$usc$ & 5020.17 & $usb$ & 15032.30 & $usc$ & \phantom{0}3761.48 & $dsc$ & 3760.88 & $usb$ & 10394.83 & $dsb$ & 10399.08 \\
$ucb$ & 5089.70 & $ucb$ & 14972.85 & $ucb$ & \phantom{0}3803.80 & $dcb$ & 3805.60 & $ucb$ & 10432.20 & $dcb$ & 10418.20 \\
$dsc$ & 5016.97 & $dsb$ & 14978.20 & & & & \\
$dbc$ & 5092.40 & $dcb$ & 15011.30 & & & & \\
\cline{1-2} \cline{3-4} \cline{5-6} \cline{7-8} \cline{9-10} \cline{11-12}
Average & 5061.89 & Average & 15006.29 & Average & \phantom{0}3789.83 & Average & 3790.19 & Average & 10420.67 & Average & 10417.58\\
$\sigma$ & \phantom{00}39.57 & $\sigma$ & \phantom{000}29.75 & $\sigma$ & \phantom{000}24.55 & $\sigma$ & \phantom{00}25.39 & $\sigma$ & \phantom{000}22.42 & $\sigma$ & \phantom{000}18.19 \\
\hline
\hline
Multiplet & $\Omega^{*+}_\mathrm{cc}$ & Multiplet & $\Omega^{*-}_\mathrm{b}$ & Multiplet & $\Omega^{*-}_\mathrm{bb}$ & Multiplet & $\Omega^{*+}_\mathrm{ccb}$ & Multiplet & $\Omega^{*0}_\mathrm{cbb}$ & Multiplet & $\Sigma^{*0}_\mathrm{b}$ \\
\cline{1-2} \cline{3-4} \cline{5-6} \cline{7-8} \cline{9-10} \cline{11-12}
$usc$ & 3901.93 & $usb$ & \phantom{0}6104.77 & $usb$ & 10538.81 & $ucb$ & 8403.90 & $ucd$ & 11718.10 & $udb$ & \phantom{0}5833.60 \\
$dsc$ & 3900.32 & $dsb$ & \phantom{0}6107.73 & $dsb$ & 10542.97 & $dcb$ & 8398.70 & $dcb$ & 11705.00 \\
\cline{1-2} \cline{3-4} \cline{5-6} \cline{7-8} \cline{9-10} \cline{11-12}
Average & 3901.13 & Average & \phantom{0}6106.25 & Average & 10540.89 & Average & 8401.30 & Average & 11711.55 & Average & \phantom{0}5833.60 \\
$\sigma$ & \phantom{000}1.14 & $\sigma$ & \phantom{0000}2.09 & $\sigma$ & \phantom{0000}2.94 & $\sigma$ & \phantom{000}3.68 & $\sigma$ & \phantom{0000}9.26 & $\sigma$ & --- \\
\hline
\hline
Multiplet & $\Xi^{*+}_\mathrm{cb}$ & Multiplet & $\Xi^{*-}_\mathrm{b}$ & Multiplet & $\Xi^{*0}_\mathrm{cb}$ & Multiplet & $\Omega^{*0}_\mathrm{cb}$\\
\cline{1-2} \cline{3-4} \cline{5-6} \cline{7-8}
$ucb$ & 7118.00 & $dsb$ & \phantom{0}5936.84 & $ucb$ & 7111.90 & $scb$ & \textbf{7241.19} \\
\cline{1-2} \cline{3-4} \cline{5-6} \cline{7-8} 
Average & 7118.00 & Average & \phantom{0}5936.84 & Average & 7111.90 & Average & \textbf{7241.19} \\
$\sigma$ & --- & $\sigma$ & --- & $\sigma$ & --- & $\sigma$ & --- \\
\end{tabular}
\end{ruledtabular}
\end{table*}

\newpage
\begin{table*}[!]
\caption{\label{tab-5}$Q^{ijk}_\mathrm{abs}$ and $\tilde{Q}^{ijk}_\mathrm{rel}$ estimates vs. direct determinations\footnote{Plain values were determined using only the PDG baryon masses, while values in bold were estimated by completing the multiplet with the average baryon masses from Table~\protect\ref{tab-PREDICT-O} and Table~\protect\ref{tab-PREDICT}.}}
\renewcommand{\arraystretch}{1.4}
\begin{ruledtabular}
\begin{tabular}{crrrrrrc}
\multirow{2}{*}{$ijk$} & \multicolumn{3}{c}{$Q^{ijk}_\mathrm{abs}$ (MeV)} & \multicolumn{3}{c}{${\tilde Q}^{ijk}_\mathrm{rel}$} & \multirow{2}{*}{Note} \\ 
\cline{2-4} \cline{5-7}
{} & $m_i$-based\footnotemark[2] & Octet-based & Decuplet-based & $m_i$-based\footnotemark[2] & Octet-based & Decuplet-based \\ 
\hline
$uds$ & 2.5 & 4.05 & 0.80 & $9.11 \times10^{-3}$ & $1.07 \times 10^{-2}$ & $1.81 \times 10^{-3}$ & \footnotemark[3] \\ 
$udc$ & 2.5 & 0.16 & $-0.13$ & $6.55 \times10^{-4}$ & $5.76 \times 10^{-5}$ & $-3.37 \times 10^{-5}$ & \footnotemark[3] \\ 
$udb$ & 2.5 & 1.94 & 0.43 & $1.79 \times10^{-4}$ & $2.05 \times 10^{-4}$ & $3.11 \times 10^{-5}$ & \footnotemark[3] \\ 
$usc$ & 92.7 & 121.59 & 140.45 & $2.52 \times10^{-2}$ & $4.89 \times 10^{-2}$ & $3.94 \times 10^{-2}$ & \\ 
$usb$ & 92.7 & 125.49 & 143.98 & $6.72 \times10^{-3}$ & $1.37 \times 10^{-2}$ & $1.07 \times 10^{-2}$ & \\ 
$ucb$ & 1272.7 & \textbf{1242.88} & 1285.90 & $1.06 \times10^{-1}$ & $\mathbf{1.57 \times 10^{-1}}$ & $1.08 \times 10^{-1}$ & \\ 
$dsc$ & 90.2 & 121.44 & 140.24 & $2.45 \times10^{-2}$ & $4.91 \times 10^{-2}$ & $3.94 \times 10^{-2}$ & \\ 
$dsb$ & 90.2 & 127.05 & 143.89 & $6.53 \times10^{-3}$ & $1.39 \times 10^{-2}$ & $1.06 \times 10^{-2}$ & \\ 
$dcb$ & 1270.2 & \textbf{1241.97} & 1286.80 & $1.06 \times10^{-2}$ & $\mathbf{1.57 \times 10^{-2}}$ & $1.09 \times 10^{-1}$ & \\ 
$scb$ & 1180.0 & \textbf{1011.37} & \textbf{1137.02} & $9.92 \times10^{-2}$ & $\mathbf{1.42 \times 10^{-2}}$ & $\mathbf{9.75 \times 10^{-2}}$ & 
\end{tabular}
\footnotetext[2]{$Q_\mathrm{abs}^{ijk}$ estimates are based on the $\overline{\mathrm{MS}}$ quark masses from \protect\cite{PDG2012} and should only be used for order-of-magnitude considerations.} \\ 
\footnotetext[3]{The measured $Q^{ijk}_\mathrm{abs}$ are sensitive to electromagnetic interactions which we neglected in our generalized GMO formalism. The $udk$ multiplets values will be particularly sensitive to EM corrections.}
\end{ruledtabular}
\end{table*}

\end{document}